\RequirePackage{fix-cm}

\documentclass[aps,prl,superscriptaddress,showpacs,onecolumn,floatfix,amsmath,amssymb]{revtex4-2}
\usepackage{graphicx}
\usepackage{upgreek}
\usepackage{hyperref}
\usepackage{amsmath}

\usepackage{xcolor}

\newcommand{\Fnp}{\ensuremath{F_2^n/F_2^p}}
\newcommand{\HeT}{\ensuremath{^3}\text{He/}\ensuremath{^3}\text{H}}
\newcommand{\HeD}{\ensuremath{^3}\text{He/}\ensuremath{^2}\text{H}}
\newcommand{\TD}{\ensuremath{^3}\text{H/}\ensuremath{^2}\text{H}}

\begin{document}

\title{The Jefferson Lab tritium program of nucleon and nuclear structure measurements}


\author{J. Arrington}  
\affiliation{Lawrence Berkeley National Laboratory, Berkeley, California 94720, USA}
\author{R. Cruz-Torres}  
\affiliation{Lawrence Berkeley National Laboratory, Berkeley, California 94720, USA}
\author{T. J. Hague}  
\affiliation{Lawrence Berkeley National Laboratory, Berkeley, California 94720, USA}
\author{L. Kurbani}  
\affiliation{University of New Hampshire, Durham, New Hampshire 03824, USA}
\author{S. Li}  
\affiliation{Lawrence Berkeley National Laboratory, Berkeley, California 94720, USA}
\author{D. Meekins}
\affiliation{Thomas Jefferson National Accelerator Facility, Newport News, Virginia 23606, USA}
\author{N. Santiesteban}  
\affiliation{University of New Hampshire, Durham, New Hampshire 03824, USA}





\begin{abstract}

A series of experiments were performed in Hall A of Jefferson Lab in 2018 that used a novel tritium and helium-3 target system.  These experiments took advantage of the isospin symmetry of these mirror nuclei to make precise measurements of isospin dependence in both nucleon and nuclear structure. We summarize here the design and properties of these cells, the physics measurements that have been published, and results currently under analysis from this program.

\end{abstract}
\maketitle

\section{Introduction} 

\subsection{Tritium Program Overview}\label{sec:intro}

While protons and neutrons represent the primary building blocks of matter and the main source of mass in the visible universe, the structure of the proton is significantly better understood due to the lack of free-neutron targets for scattering measurements. As such, deuterons are frequently used to make electron-scattering measurements on the neutron. However, this involves correcting for the e-p cross section, which typically dominates, as well as correcting for nuclear effects, e.g. binding and Fermi motion, that make the proton and neutron in the nucleus differ from free, stationary proton and neutron targets. These corrections are typically quite small, but still yield the dominant source of uncertainty in extractions of the neutron structure function~\cite{Arrington:2011qt} and parton distribution functions~\cite{Accardi:2011fa} at large $x$. They also contribute to the uncertainty in the extraction of neutron form factors from quasi-elastic (QE) scattering from neutrons in the deuteron~\cite{Arrington:2006zm}. 

An alternative technique is to compare scattering from the mirror nuclei $^3$H and $^3$He. While the nuclear effects are larger in these nuclei, it is expected that they will be nearly identical~\cite{Afnan:2003vh} and thus the comparison of $^3$H and $^3$He should yield a theoretically cleaner extraction of neutron structure. This allows for measurements of neutron structure, both structure functions and elastic form factors, using simple inclusive scattering and requiring minimal corrections for nuclear effects.

In addition to using the isospin symmetry between $^3$H and $^3$He to measure the neutron structure, we can also make precision measurements of nuclear structure in these well-understood few-body nuclei. By measuring single proton knockout, we can compare the proton distributions in $^3$He and $^3$H, which by isospin symmetry provide a comparison of the proton and neutron distributions in $^3$H.  At low momenta, these measurements can be compared to \textit{ab initio} calculations of the proton and neutron distributions in $^3$H~\cite{Wiringa:2013ala}, which is related to the difference in the distributions for the singly and doubly occupied shell. At high momenta, the data can test the prediction that short-range correlations (SRCs) dominate the nuclear momentum distribution above the Fermi momentum, and that these SRCs are almost completely dominated by np-SRC configurations~\cite{Arrington:2022rev}. Inclusive scattering at large momentum transfer but small energy transfer can also isolate scattering from high-momentum nucleons, providing excellent statistics, and allowing for a much more precise extraction of the relative contribution of np- and pp-SRCs in light nuclei, while avoiding the large corrections associated with rescattering of the struck nucleon and the spectator nucleon from the SRC.

Finally, the availability of the $^3$H target also provided a unique opportunity to study charge symmetry breaking in the $\Lambda$-N interaction~\cite{E12-17-003}. While a $\Lambda$nn bound state has been ruled out by theoretical analyses with the constraints of existing $\Lambda$p data, such a physical resonance could exist if the $\Lambda$-n interaction is as little as $\sim$5\% stronger than the $\Lambda$-p interaction. As the $\Lambda$-n interaction cannot be studied directly, the study of a $\Lambda$nn physical resonance pole is a powerful tool to place constraints on the strength of $\Lambda$-N charge symmetry breaking. These results saw evidence of a $\Lambda$nn resonance, but without sufficient statistical significance to definitively claim the existence of the resonance~\cite{HallA:2022qqj}. From this, the cross section to produce the potential resonance was calculated for constraining models of the $\Lambda$-n interaction~\cite{Suzuki:2021pid}. While this experiment required the availability of a tritium target, it is not focused on conventional nuclear structure or on a comparison of scattering from $^3$H and $^3$He, and as such we do not discuss this measurement in further detail below. 

The primary challenge to these studies is obtaining a well-understood $^3$H target of sufficient density to allow for precision comparisons to $^3$He.  Tritium targets have been used at electron beam facilities at the Stanford High Energy Physics Laboratory~\cite{Hughes:1966zza}, MIT-Bates~\cite{Beck:1984qhv,Beck:1989bi}, and at Saclay~\cite{Juster:1985sd}, most recently in 1985. The basic target parameters are compared to the target designed for the present experiment in Ref.~\cite{target_NIM}. While the luminosity of this target is not higher than the previous designs, the activity is much lower and the target thickness can be well measured.

In 2018, a series of measurements ran using this new target system, allowing for measurements from identical low-pressure gas target cells of $^1$H, $^2$H, $^3$H, and $^3$He.  While the combination of the Jefferson Lab beam and these cells were of comparable luminosity to previous measurements, the combination of high beam energies and relatively large acceptance spectrometers allowed for significant improvements compared to previous measurements utilizing tritium targets.  We describe the target system, including the special considerations involved in making precision measurements on sealed gas cells, in the following section. We then summarize the result and ongoing analyses from a series of measurements studying neutron and nuclear structure from the 2018 experimental campaign in Hall A.

This new target, described in detail in the following section, was coupled with the two High-Resolution Spectrometers (HRSs) in Hall A at Jefferson Lab. A general description of the beamline instrumentation and HRS spectrometer and detector packages can be found in Ref.~\cite{Alcorn:2004sb}. Additional detail is included with each physics topic, summarizing key aspects of the analysis procedures with particular emphasis on corrections that are especially important for the individual measurement.

\subsection{Tritium Target Design}\label{sec:target}

\begin{figure}[!htb]
    \centering
    \includegraphics[width=0.51\textwidth,trim={20mm 0mm 220mm 0mm},clip]{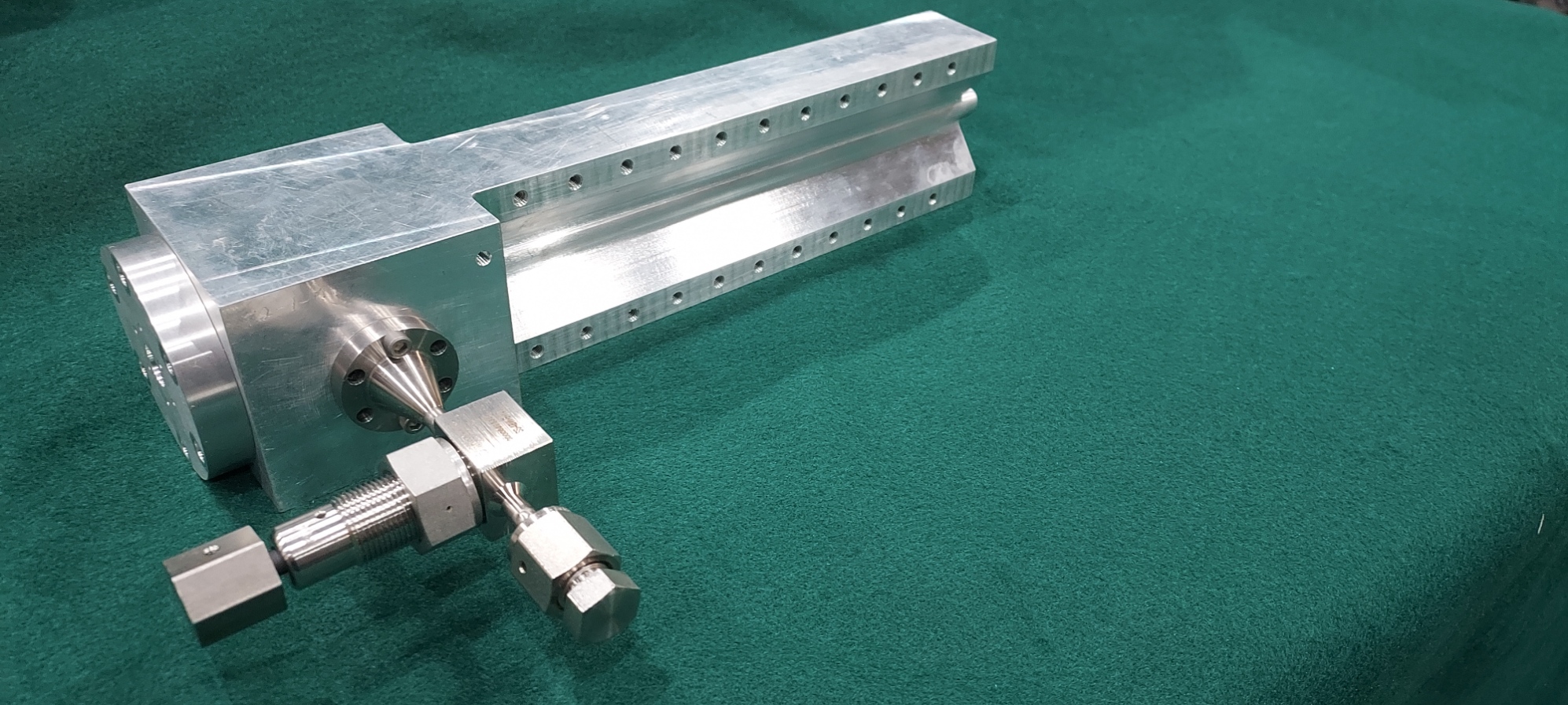}
    \includegraphics[width=0.46\textwidth,trim={140mm 0mm 160mm 0mm},clip]{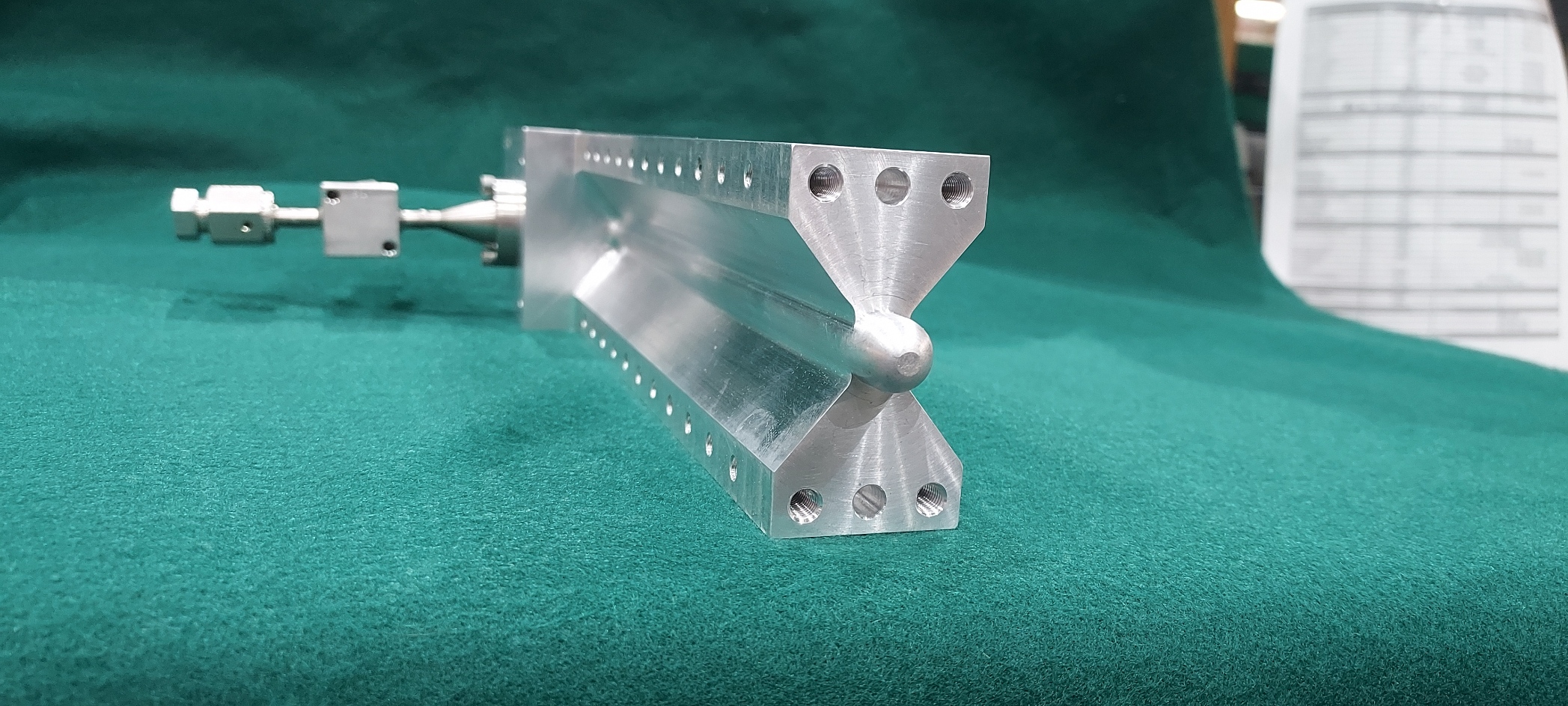}
    \caption{Example of a target cell assembly. In the left view, the entrance window assembly can be seen on the far left of the figure. The electron beam passed through the cell and exited through a hemispherical shaped window as shown on the right view.}
    \label{fig:T2-cell}
\end{figure}

The standard Hall A Cryogenic Target was significantly altered to provide a system that would safely contain tritium gas in an electron beam environment while still meeting the performance needs of the experimental program. The design was, in part, based on a proposed design presented in~\cite{target_NIM}. The system that was developed incorporated a modular sealed gas cell design that allowed the cells to be filled at room temperature in convenient locations prior to installation. Each cell assembly consisted of three main components: the main body, the entrance window, and the fill tube with a valve which were bolted together using aluminum 1100-O seals. The entrance window and main body were each fabricated from a single plate of 7075-T851 aluminum and had a design pressure of 3.44 MPa. The entrance window assembly was machined as a cap to a re-entrant tube/flange in the rough shape of a top hat. The welded stainless steel fill tube assembly included a valve altered for tritium use. The cell assembly, shown in Figure~\ref{fig:T2-cell} has a tubular shaped gas volume with a diameter of 1.27 cm and a length of 25 cm. The additional material in the shape of an hourglass, provided additional strength and material for conducting heat from beam exit of the cell. The hourglass shape ensured that this additional material was well removed from the acceptance of the spectrometers while leaving thin walls on the sides where the spectrometers detect the scattered electrons. The walls of the cell were 0.5 mm thick and and the beam entrance (not visible in Figure~\ref{fig:T2-cell}) and exit (the domed shape shown in Figure~\ref{fig:T2-cell} right view) to the cell were 0.25 mm thick.

\begin{figure}[!htb]
    \centering
    \includegraphics[width=0.48\textwidth,trim={20mm 2mm 30mm 28mm}, clip]{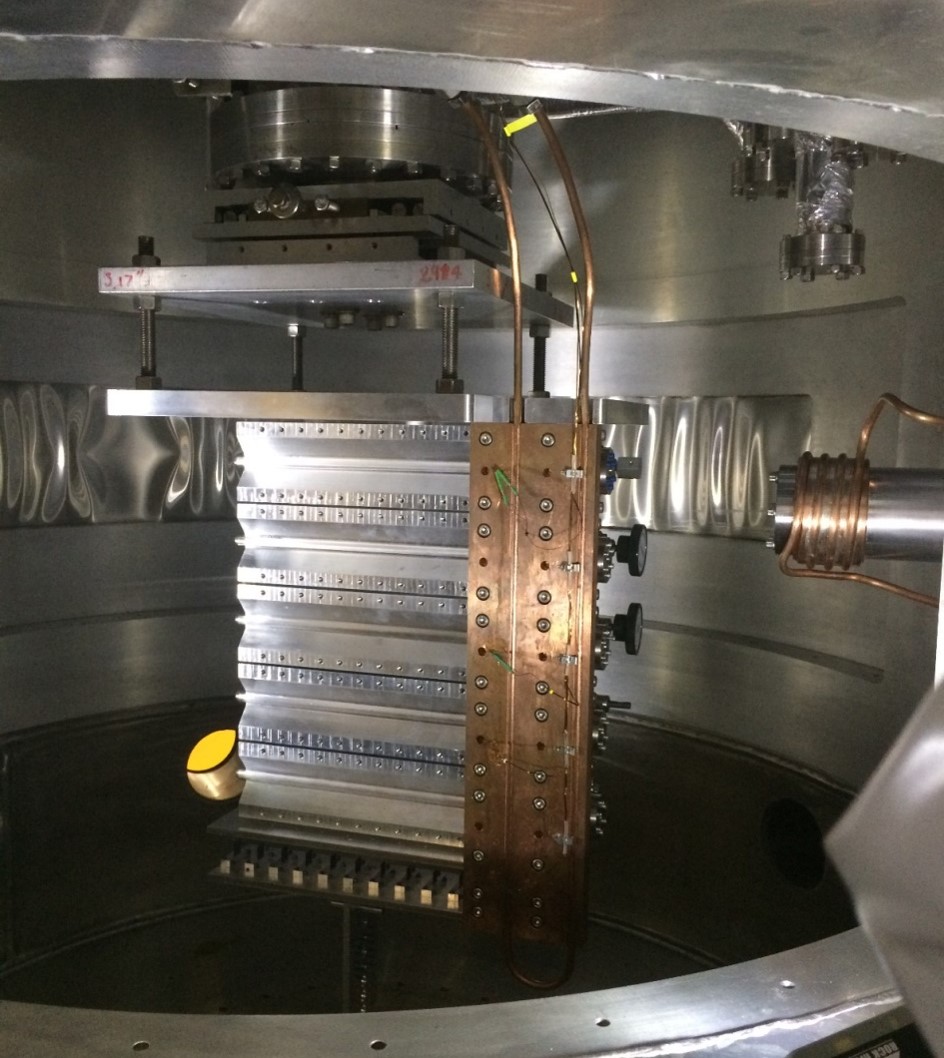}
    \caption{Target assembly in the scattering chamber.  On top are five cells, followed by a series of foils along the beam direction for acceptance studies, and at the bottom is a ladder of various solid target foils.}
    \label{fig:tgt-stack}
\end{figure}

Four gas cells were installed in the target system; each filled with hydrogen, deuterium, tritium, and helium-3 with areal densities 70.8, 142.2, 84.8 and 53.2~mg/cm$^2$ respectively. An identical empty cell, used for background studies, was also installed. Additional solid targets were installed for calibration measurements. The assembly in the scattering chamber is shown in Figure~\ref{fig:tgt-stack}. The modular cell design allowed the hydrogen, deuterium, and helium-3 individual cells to be filled in a clean room environment. It also allowed for the the tritium cells to be filled offsite at Savannah River Tritium Enterprises, a Department of Energy Facility located on the Savannah River Site. In this way, actual tritium gas handling at JLab was avoided and instead the tritium target was always handled as a sealed cell.

During operations, the electron beam induced a significant thermal load on the aluminum cell walls. This heat load has large and frequent changes associated with the electron beam turning on and off due to beam trips and experimental configuration changes. To ensure that the thermal load remained within acceptable limits, the target cell main bodies were cooled to 40K. The electron beam was limited to 22.5~$\mu$A and rastered to a $2\times2$~mm$^2$ square. The depth and number of the thermal cycles also required a second tritium cell to be prepared and installed after the first run period ended to avoid unacceptable fatigue in the wall material. The target gas, also heated by the beam, quickly reaches an equilibrium state within the cell with a reduced gas density along the beam path. A detailed study~\cite{Santiesteban:2018qwi} shows that the tritium, deuterium, and helium-3 areal densities as seen by the beam decreased by 9.7\%, 9.0\%, and 6.2\%, respectively, at 22.5~$\mu$A. Detailed descriptions of the beam line instrumentation can be found in Ref.~\cite{Alcorn:2004sb}.

The tritium in the target cell beta decays into helium-3 with a half-life of 12.3 years. Thus over the period of one year the amount of tritium would be reduced by~5.5\% with a corresponding $^3$He contamination increase. The first target cell was filled and installed in 2017 yielding an average 4\% $^3$H density reduction over the initial run period. For the second cell, in the corresponding run period, this reduction was 1\%.  The resulting $^3$He contamination in the $^3$H data was removed using $^3$He data taken at the same settings. The quasi-elastic data taken on the second tritium cell showed a narrow peak at Bjorken $x=1$, corresponding to elastic scattering from hydrogen. The best hypothesis for this hydrogen contamination is that a small amount of water was introduced when the cell was filled. Water in the cell will freeze to the walls and not contribute to the scattering. But the hydrogen in the water can exchange with tritium gas by the H$_2$O + T$_2$ $\to$ 2HTO + H$_2$ and H$_2$ + T$_2$ $\to$ 2HT reaction. The observed hydrogen contamination requires 4.1\% of tritium gas in the tritium cell to have exchanged with hydrogen in the water to form HTO, which freezes on the target wall and so is removed from the effective target thickness. To correct for this, we used the hydrogen cell data to subtract the hydrogen contribution to the tritium cell and reduce the effective tritium target thickness by 4.1$\pm$0.2\%. This correction was applied for all of the (e,e') data analysis that included data taken on the second tritium cell.

\section{The Tritium Physics Program}\label{sec:physics_overview}

The following sections describe the analysis and results of the main proposals focusing on neutron and nuclear structure from comparisons of electron scattering on $^3$H and $^3$He~\cite{E12-10-103,E12-11-112,E12-14-011,E12-10-003}. We summarize the measurements, analysis procedures, and published results for these experiments, and provide additional information on results that are still under analysis. As noted in the introduction, we do not present the tritium hypernuclear experiment~\cite{E12-17-003}, which used the tritium target to look for A=3 hypernuclei and study the $\Lambda$-n interaction~\cite{HallA:2022qqj}. Several of these results were highlighted in the White Paper produced from the QCD Town Hall~\cite{Achenbach:2023pba}, and the program was successful enough that additional experiments have been proposed~\cite{E12-20-005,E12-21-004} or are planned~\cite{Arrington:2021alx} to extend the tritium physics program at Jefferson Lab.

\subsection{MARATHON: DIS to extract neutron pdfs and the A=3 EMC effect}\label{sec:MARATHON}
MARATHON (experiment E12-10-103)~\cite{E12-10-103} studied parton structure through deep-inelastic scattering (DIS) on $^3$H and $^3$He. In the absence of a free neutron target, MARATHON sought to utilize the expectation that nuclear effects are nearly identical in $^3$H and $^3$He to study the neutron to proton structure function ratio, \Fnp{}~\cite{Afnan:2003vh}. This expectation can be quantified, in a model-dependent manner, by forming an ``EMC-type'' ratio for each target. These ratios represent the deviation of nuclear structure from a mere sum of the constituent nucleons such that, 
\begin{align}
     R_h = \frac{F_2^{^3\text{He}}}{2F_2^p+F_2^n}~~~~, && R_t = \frac{F_2^{^3\text{H}}}{F_2^p+2F_2^n}.
     \label{eq:emc-type}
\end{align}
A ``super ratio'' of these can be formed to quantify the difference in nuclear effects of the two nuclei,
\begin{equation}
     \mathcal{R} = \frac{R_h}{R_t},
     \label{eq:superR}
\end{equation}
where a value of $\mathcal{R}(x)=1$ represents the two nuclei having identical nuclear effects. From Eqs.~\ref{eq:emc-type} and~\ref{eq:superR}, we can extract the nucleon structure function ratio in terms of the $^3$He/$^3$H ratio and $\mathcal{R}$,
\begin{equation}
     \frac{F_2^n}{F_2^p} = \frac{\frac{F_2^{^3\text{He}}}{F_2^{^3\text{H}}} - 2\mathcal{R}}{\mathcal{R} - \frac{F_2^{^3\text{He}}}{F_2^{^3\text{H}}}}.
\end{equation}
MARATHON used the assumption that the ratio of structure functions of nuclei is equal to the ratio of the cross sections of the nuclei.
This equality is rooted in the assumption that $R=\sigma_L/\sigma_T$, the ratio of the cross sections for absorbing longitudinal and transverse helicity virtual photons, is identical for all nuclear targets~\cite{Geesaman:1995yd,E140X:1995ims}.
While this assumption is still under further investigation~\cite{E12-14-002}, it is standard practice in nucleon structure experiments as all current measurements find that the assumption holds within uncertainties.

\begin{figure}
    \centering
    \includegraphics[width=0.65\textwidth]{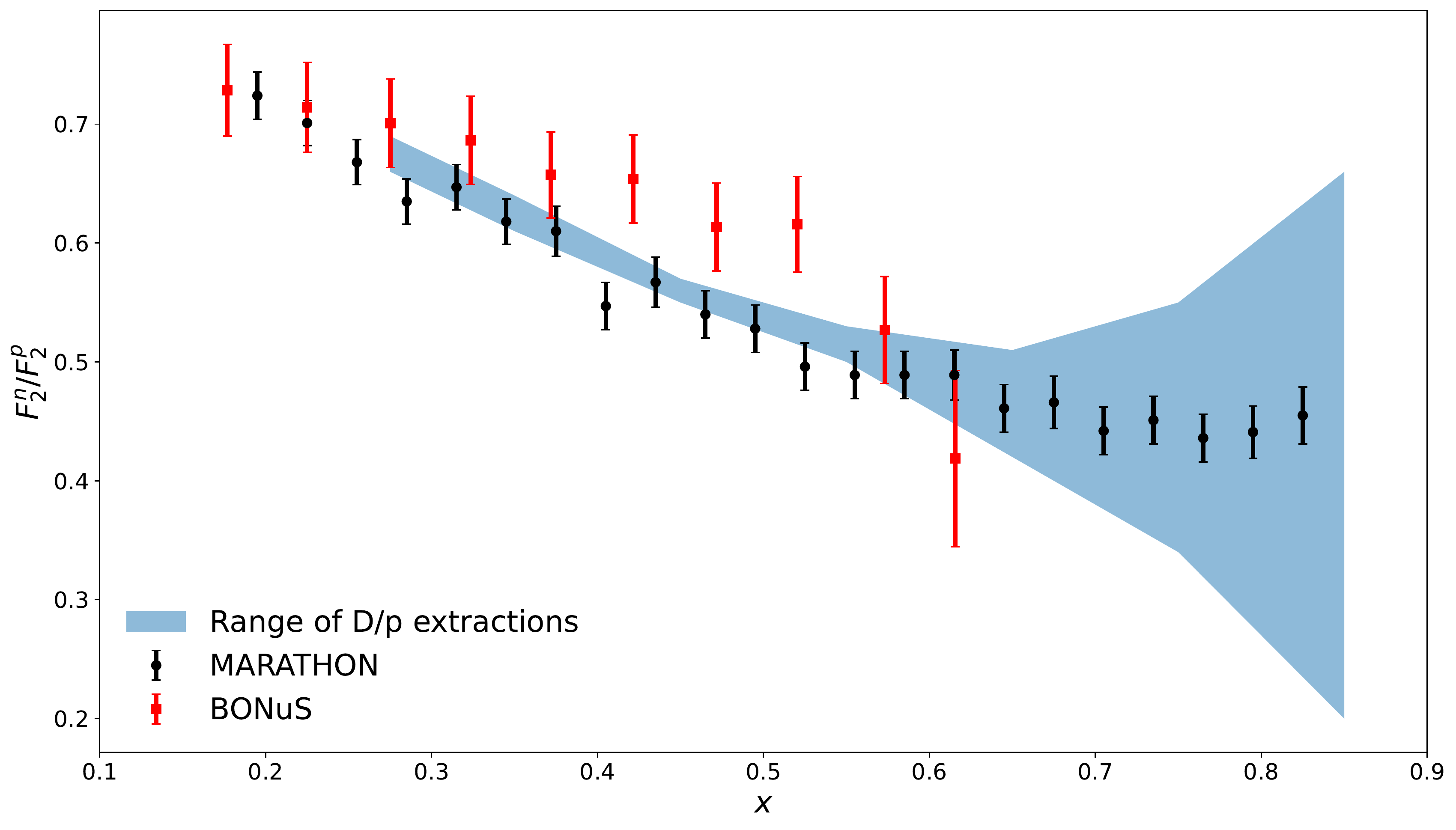}
    \caption{\Fnp{} structure function ratio extracted from MARATHON~\cite{Abrams:2021xum} \HeT{} data (including the 2.5\% normalization shift to the cross section ratio from~\cite{Abrams:2021xum}). Also shown are the BONuS results~\cite{CLAS:2014jvt} and a band indicating the range of \Fnp{} values extracted from $^2$H/$^1$H data~\cite{Arrington:2011qt}.}
    \label{fig:marathon_np}
\end{figure}

In 2022 the MARATHON experiment published their \Fnp{} results using this technique~\cite{Abrams:2021xum}. This result used the Kulagin-Petti (KP) model~\cite{Kulagin:2004ie,Kulagin:2010gd,Alekhin:2017fpf} to evaluate $\mathcal{R}$. The analysis additionally used the $^2$H/$^1$H ratio to extract \Fnp{} using an analogous technique, which was found to disagree with the extraction from \HeT{}. The $^2$H/$^1$H ratio was shown to be in good agreement with prior SLAC data~\cite{Bodek:1979rx} in the region where the nuclear effects are expected to be negligible ($x\approx0.3$), which was taken as a validation of the extraction from $^2$H/$^1$H. The disagreement was assumed to be associated with the normalization of the target thicknesses, and the \HeT{} normalization was modified to reproduce the $^2$H/$^1$H extractions of \Fnp{} in this region.
A normalization of 2.5\% was applied to the \HeT{} data in order to satisfy this condition, yielding the \Fnp{} extraction in Fig.~\ref{fig:marathon_np}, which shows the result alongside the \Fnp{} results from the BONuS experiment~\cite{CLAS:2014jvt} and a band indicating the range of \Fnp{} extractions from $^2$H/$^1$H data~\cite{Arrington:2011qt}. 
This extraction has been used in a bootstrap fitting analyses to extrapolate to $x\rightarrow0$ and $x\rightarrow1$ in order to provide further model constraints~\cite{Valenty:2022gqm,Ding:2022ows}.

The \HeT{} and $^2$H/$^1$H ratios can be included in any global pdf analysis that either uses a model for the nuclear effects or includes an isospin-independent nuclear effect as part of the parameterization. 
This would improve the extraction of the $d(x)/u(x)$ ratio in the analysis, especially at large $x$, while also improving the constraints on off-shell effects where they are included as part of the fit~\cite{Segarra:2021exb,Alekhin:2022tip,Alekhin:2022uwc}. 
In addition to this, the JAM collaboration showed that adding the MARATHON data to global pdf extractions makes possible a modified extraction to probe the flavor dependence of the off-shell effects~\cite{Cocuzza:2021rfn}.
These effects are typically taken to be flavor independent, with the exception of a small difference from Fermi smearing where the faster falloff of the $d$-quark contribution at large $x$ yields a slightly larger fractional enhancement in nuclei.
By allowing for different off-shell effects for $u(x)$ and $d(x)$, the comparison of deuteron and proton structure functions helps to constrain the average off-shell corrections, while the $^3$H and $^3$He data constrain the flavor dependence. 
This technique also allowed for an extraction of $\mathcal{R}(x)$ that strongly deviates from the KP model and does not predict the cancellation of nuclear effects at $x=0.3$.
The results of this analysis find that the MARATHON data are consistent with flavor-dependent off-shell effects, suggesting the existence of an isovector EMC effect.

These data also allow for the first measurement of the EMC effect in $^3$H and a new measurement of $^3$He at the same kinematics and the first measurement on $^3$He fully outside of the resonance region up to $x\approx 0.8$. Such extractions require an isoscalar correction applied to the \TD{} and \HeD{} ratios (to correct for unequal proton and neutron numbers),
\begin{equation}
    \left.\frac{F_2^A}{F_2^{^2\text{H}}}\right|_{\text{iso}} = \frac{F_2^A}{F_2^{^2\text{H}}}\cdot\frac{A\left(1 + \Fnp{}\right)}{1\left(Z + \left(A-Z\right)\Fnp{}\right)},
\end{equation}
which makes the results dependent on the assumed \Fnp{} used in the extraction. This is a general issue with EMC effect measurements on non-isoscalar nuclei, but the $^3$H/$^2$H and $^3$He/$^2$H ratios are particularly sensitive to the isoscalar correction due to their highly asymmetric $N/Z$ ratios, and the comparison of the $^3$H and $^3$He EMC effects are even more sensitive. 

However, in addition to extracting the model-dependent EMC ratios for $^3$H and $^3$He, these data can also be used to extract the isoscalar-averaged EMC ratio, i.e. the (${}^3\text{H}+ {}^3\text{He}$)/($2\cdot {}^2\text{H}$) cross section ratio. This provides a model-independent extraction for the EMC effect of a hypothetical isoscalar A=3 nucleus, allowing us to examine the size of the EMC effect in light nuclei without the model dependence associated with the isoscalar corrections. A manuscript describing the analysis and results of the EMC ratios is currently in preparation.

\subsection{Quasi-elastic proton knockout from A=3 nuclei}

Experiment E12-14-011~\cite{E12-14-011} was carried out to constrain models of the nucleon-nucleon ($NN$) interaction and reaction mechanisms in light nuclei with high precision, mainly in the high-nucleon-momentum regime dominated by short-range correlations (SRC). In this context, low and high is relative to the nuclear Fermi momentum, $k_{\rm F}$ ($\approx 250 \ {\rm MeV}$ for a medium-size nucleus). In these measurements, an electron scatters from a proton, knocking it out of the nuclear system. The scattered electron and struck nucleon are measured in coincidence (i.e. both detected in the same event).

In the plane-wave impulse approximation (PWIA) for quasi-elastic (QE) scattering, an electron $(E_e,\overrightarrow p_e)$ scatters into a final state $(E_{e'},\overrightarrow p_{e'})$ off a single proton $(E_p,\overrightarrow p_p)$ in the nucleus via the exchange of a single virtual photon $(\omega=E_e-E_{e'},\overrightarrow q = \overrightarrow p_e-\overrightarrow p_{e'})$. The proton is ejected from the nucleus without any subsequent rescattering. Reconstructing the kinematics of both the scattered electron and knocked-out proton allows us to determine the missing momentum and missing energy, defined as:
\begin{align}
    \overrightarrow p_{\rm miss} = \overrightarrow p_p - \overrightarrow q, && E_{\rm miss} = \omega - T_p -T_{\rm A-1},
\end{align}
where $T_p$ and $T_{\rm A-1}$ are the kinetic energies of the knocked-out proton and recoiling A-1 system, respectively. The kinetic energy of the A-1 system is calculated assuming that the recoiling system is a bound A-1 nucleus, with momentum equal to but opposite of $p_{\rm miss}$.
In the PWIA, the differential cross section for such a process ($\sigma_{\rm PWIA}$) is proportional to the spectral function $S_{\rm A}(|\overrightarrow p_i|,E_i)$:

\begin{equation}
    \frac{{\rm d}^6 \sigma_{\rm PWIA}}{{\rm d} \omega {\rm d} E_p {\rm d}\Omega_e {\rm d}\Omega_p} = K \, \sigma_{ep} \, S_{\rm A}(|\overrightarrow p_i|,E_i),
    \label{eq:pwia}
\end{equation}
where $\Omega_i$ corresponds to the electron and proton solid angles, $K$ is a kinematic factor, and $\sigma_{ep}$ is the elementary cross section for electrons scattering off bound protons~\cite{DeForest:1983ahx}.
The nuclear spectral function encodes the probability of finding a nucleon inside nucleus A with momentum $\overrightarrow p_i$ and separation energy $E_i$~\cite{Lapikas:1993uwd}. Integrating the spectral function over $E_i$ results in the single-nucleon momentum distribution: $n_{\rm A}(|\overrightarrow p_i|) = \int S_{\rm A}(|\overrightarrow p_i|,E_i) \, dE_i$. Both of these quantities are commonly used to describe the internal structure of nuclei.
In the PWIA, the missing momentum and missing energy are equal to the initial nucleon momentum and separation energy ($\overrightarrow p_i = \overrightarrow p_{\rm miss}, E_i = E_{\rm miss}$).
While spectral functions and momentum distributions are not measurable observables, a careful event selection can provide sensitivity to their properties.

Previous measurements~\cite{JeffersonLabHallA:2004tej} found large disagreements between the measured $^3$He$(e,e'p)$ cross section and PWIA calculations for quasielastic scattering in the high-$p_{\rm miss}$ regime. This was attributed to competing non-QE reaction mechanisms such as meson-exchange currents (MEC), isobar configurations (IC), and rescattering of the struck nucleon (final-state interactions (FSI)), which result in the same final state as the PWIA channel. Such competing reaction mechanisms can be suppressed by judicious choice of kinematics, thus regaining sensitivity to the nuclear ground state. MEC and IC are suppressed for $Q^2>1.5$~GeV$^2$ and $x > 1$~\cite{Sargsian:2001ax,Sargsian:2002wc}. Additionally, FSI effects at high-$Q^2$can be reduced by further constraining the angle between $- \overrightarrow p_{\rm miss}$ and $\overrightarrow q$ to be $\theta_{rq} < 40^{\circ}$, where FSI were predicted~\cite{Frankfurt:1996xx,Laget:2004sm,Jeschonnek:2008zg} and observed~\cite{CLAS:2007tee,Sargsian:2009hf,HallA:2011gjn,Arrington:2011xs} to be highly suppressed. Residual FSI effects are expected to partially cancel in the $^3$He/$^3$H cross section ratio given that these nuclei have the same number of nucleons.

The kinematics and event-selection criteria of the E12-14-011 experiment were designed to minimize non-QE reactions mechanisms.
Data were taken with a 20~$\mu$A electron beam with an incident energy of $E_e = 4.326$~GeV.
Scattered electrons were measured in the left High Resolution Spectrometer (HRS), which was fixed at a central momentum and angle $(p_0,\theta_0)$ = (3.543~GeV, 20.88$^{\circ}$). This corresponds to the central values $Q^2=2.0$~GeV$^2$ and $x = 1.4$.
Knocked-out protons were measured in the right HRS in two different configurations corresponding to a central momentum and angle $(p_0,\theta_0)$ = (1.481~GeV, 48.82$^{\circ}$) and (1.246~GeV, 58.50$^{\circ}$), covering $p_{\rm miss}$ ranges of 40--250 and 250--500~MeV, referred to as low- and high-$p_{\rm miss}$ settings.
Additionally, cuts requiring $\theta_{rq}<37.5^{\circ}$ and, in the high-$p_{\rm miss}$ setting, $x>1.3$ were used to further reduce non-QE competing mechanisms.
For additional event-selection criteria and information on particle-identification, corrections, and systematic uncertainties, see Refs.~\cite{JeffersonLabHallATritium:2019xlj,JeffersonLabHallATritium:2020mha}.

\begin{figure}[!htb]
    \centering
    \includegraphics[width=0.7\textwidth]{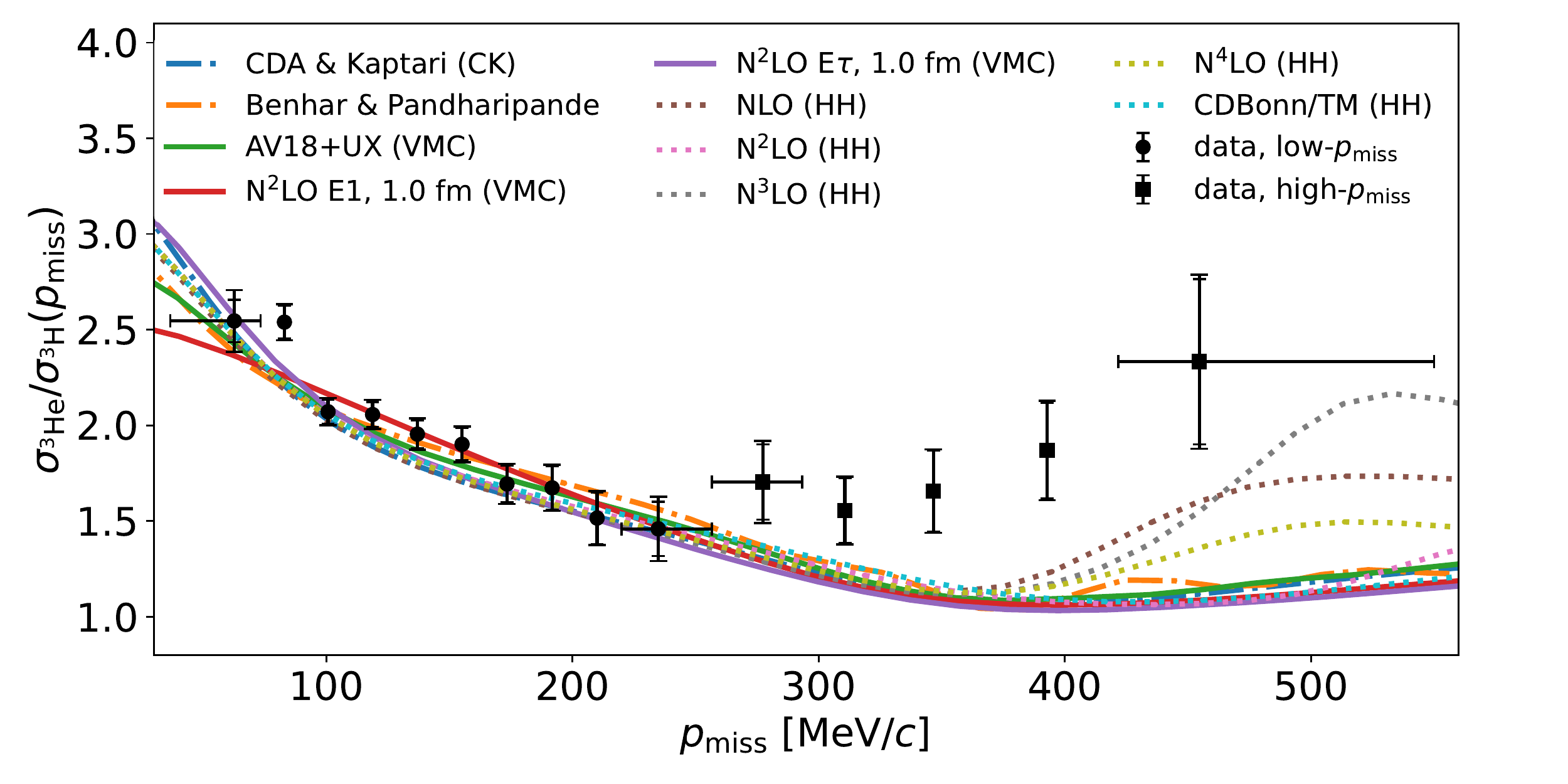}
    \caption{
    Experimental extraction of the $\sigma_{^3{\rm He}(e,e'p)}/\sigma_{^3{\rm H}(e,e'p)}$ QE cross-section ratio as a function of $p_{\rm miss}$.
    Figure adapted from Ref.~\cite{JeffersonLabHallATritium:2019xlj}.}
    \label{fig:eep1}
\end{figure}

The first result of the measurement was the $\sigma_{^3{\rm He}(e,e'p)}/\sigma_{^3{\rm H}(e,e'p)}$ cross section ratio as a function of $p_{\rm miss}$~\cite{JeffersonLabHallATritium:2019xlj}, shown in Fig.~\ref{fig:eep1}. In the ratio, systematic detector and FSI effects partially cancel, leading to a more precise result. The black circles and squares correspond to the low- and high-$p_{\rm miss}$ kinematics, respectively. The inner (outer) error bars represent statistical (total) uncertainties.

Given the selected kinematics, the measured data are expected to be sensitive to single-nucleon momentum distribution ratios. Thus, the cross-section ratio was compared to theoretical predictions based on variational Monte Carlo (VMC) techniques with local interactions~\cite{Wiringa:2013ala,Lonardoni:2018sqo} (including the AV18+UX~\cite{Wiringa:1994wb} and chiral effective field theory potentials at N$^2$LO with two parametrizations of the three-body contact term~\cite{Lonardoni:2018nob,Gezerlis:2014zia,Lynn:2015jua,Lynn:2017fxg,Lonardoni:2017hgs}) and the Hyperspherical Harmonics (HH) method with non-local interactions~\cite{Kievsky:2008es,Marcucci:2018llz} (including the CDBonn/TM potential~\cite{Machleidt:2000ge,Coon:2001pv} and chiral potentials at N$^i$LO for $i=1,2,3,4$~\cite{Entem:2017gor}). Additionally, the CDA \& Kaptari~\cite{CiofidegliAtti:2004jg} and Benhar \& Pandharipande~\cite{Benhar:1993ja} A=3 spectral functions were integrated over $E_{\rm miss}$ and included in the comparison.

The theoretical predictions agree with the measurement up to $p_{\rm miss} \approx 250 \ {\rm MeV}$. At higher missing momenta, where the $\sigma_{^3{\rm He}(e,e'p)}/\sigma_{^3{\rm H}(e,e'p)}$ ratio is expected to be close to $1$ due to $np$ dominance in SRCs, they disagree by up to $20-50\%$. While the trend in the data supports a transition between the low- and high-$p_{\rm miss}$ regions, additional studies are required to explain the excess observed at high $p_{\rm miss}$. Nevertheless, this result is a significant improvement compared to the previous QE $\sigma_{^3{\rm He}(e,e'p)}$ measurement in which the data and PWIA calculations disagreed by up to a factor of 5 in the high-$p_{\rm miss}$ region~\cite{JeffersonLabHallA:2004tej}. 

\begin{figure}[!htb]
    \centering
    \includegraphics[width=0.8\textwidth]{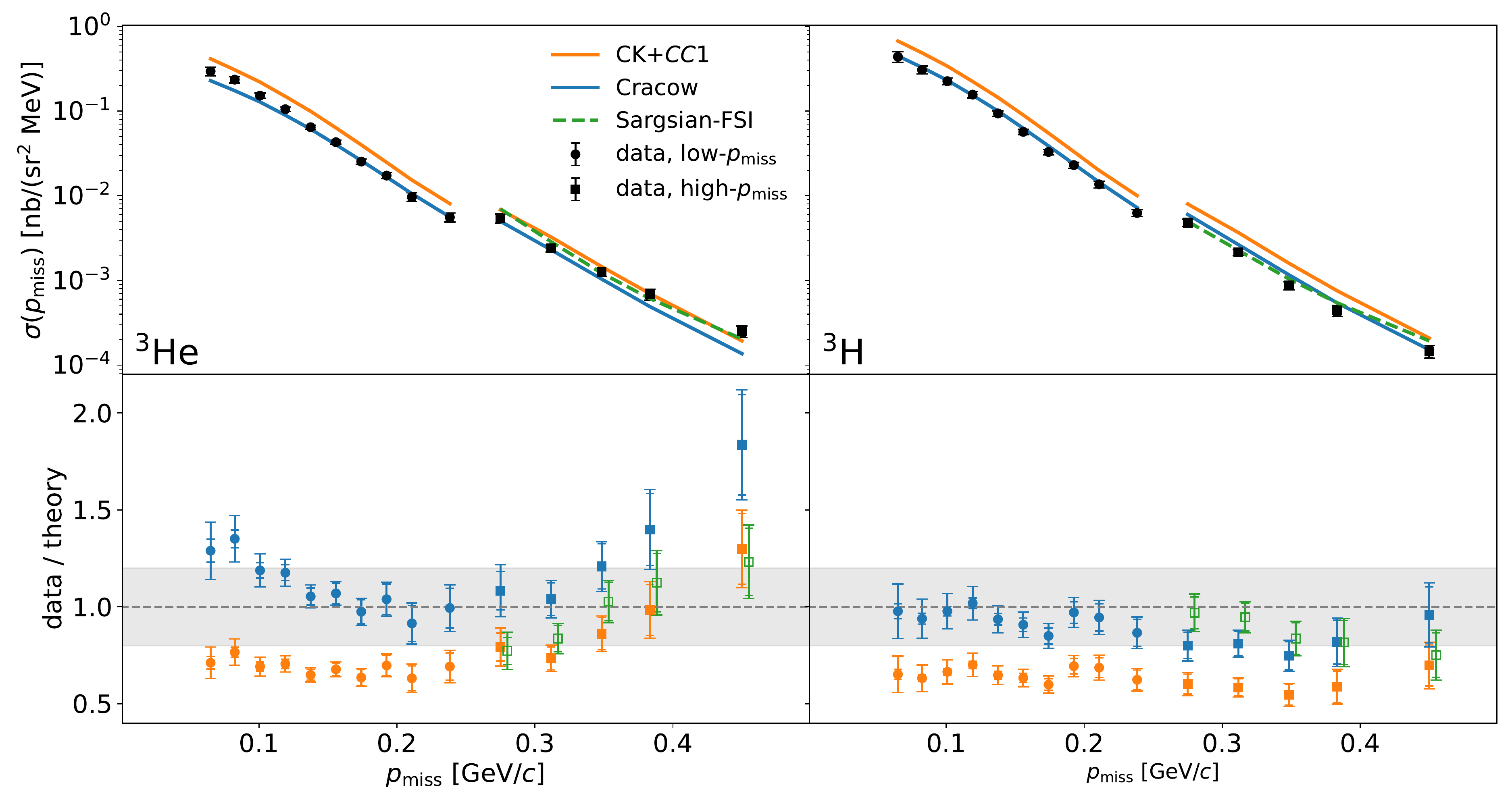}
    \caption{Absolute $^3{\rm He}(e,e'p)$ (left) and $^3{\rm H}(e,e'p)$ (right) QE cross sections as a function of $p_{\rm miss}$. Figure adapted from Ref.~\cite{JeffersonLabHallATritium:2020mha}.}
    \label{fig:eep2}
\end{figure}

From the ratio it is not possible to determine whether the discrepancy comes from $^3$He or $^3$H alone, or whether it is due to effects present in both nuclei.

To disentangle the cause of this disagreement, the E12-14-011 experiment reported the absolute $\sigma_{^3{\rm He}(e,e'p)}$ and $\sigma_{^3{\rm H}(e,e'p)}$ cross sections as a function of $p_{\rm miss}$~\cite{JeffersonLabHallATritium:2020mha}.
The results are summarized in Fig.~\ref{fig:eep2}. It is worth noting that the $^3$He/$^3$H cross section ratios, as well the cross section itself, give modest deviations from the calculations, while the cross section itself drops by almost four orders of magnitude across the measured $p_{\rm miss}$ range.

A notable difference between the $^3$He and $^3$H $(e,e'p)$ reactions is the number of final states available. The reaction in $^3$H breaks the nucleus into three nucleons: the struck proton and two neutrons (three-body breakup). The reaction in $^3$He also has a three-body breakup channel, breaking the nucleus into two protons and a neutron. In addition, this reaction can knock out a proton while leaving the unstruck proton and neutron in a bound state (deuteron). Since this channel results in two final-state particles, it is called 2-body breakup. Unlike in the ratio analysis, which includes both $^3$He channels, Ref.~\cite{JeffersonLabHallATritium:2020mha} only compared the 3-body breakup in these nuclei. This channel was isolated in the $^3$He case by requiring that $E_{\rm miss} > 8 \ {\rm MeV}$ (above the 2-body breakup threshold). The models included in this comparison correspond to 3-body breakup only.

The measured data were compared to cross-section calculations with the CD-Bonn $NN$ potential to focus on the modeling of reaction mechanisms. The PWIA {\bf Cracow} model was carried out with this potential and is based on the Faddeev formulation of the A=3 system~\cite{Golak:2005iy,Carasco:2003us}. The \textbf{CK+CC1} model is a PWIA calculation based on Eq.~\ref{eq:pwia} and uses the aforementioned CDA \& Kaptari spectral function ~\cite{CiofidegliAtti:2004jg} and the CC1 off-shell electron-proton cross section~\cite{DeForest:1983ahx}. In this case, the calculation was carried out with the AV18 potential and was subsequently scaled by the ratio of the CD-Bonn to AV18 momentum distributions.

The Cracow PWIA calculation agrees with the measured $^3$H ($^3$He) result at the $20\%$ level in the whole ($[100,350]$ MeV/$c$) $p_{\rm miss}$ range. For $^3$He at high $p_{\rm miss}$, they disagree by about 60\%. The CK+CC1 calculation overestimates the measured $^3$H ($^3$He) result by about 60\% in the whole (low-) $p_{\rm miss}$ range. For $^3$He at high $p_{\rm miss}$, this calculation agrees with the data within 20\%. These results are consistent with the disagreement found in Ref.~\cite{JeffersonLabHallATritium:2019xlj} being caused both by a deficit in the $^3$H cross section and an excess in the $^3$He case, which could arise from charge-exchange FSI, as discussed below. 

To explore the residual disagreement in the high-$p_{\rm miss}$ region, the data were compared to the model by {\bf Sargsian}~\cite{code:sargsian_fsi} which can include FSI of the struck nucleon modeled using the generalized Eikonal approximation~\cite{Sargsian:2004tz,Sargsian:2005ru}. Without FSI, this model is consistent with the PWIA calculations. The inclusion of FSI of the struck nucleon further improves the agreement with the data. There is a residual trend, with the $^3$H ($^3$He) decreasing (increasing) with $p_{\rm miss}$. A hypothesis that explains this trend is single-charge exchange (SCX) in which the struck proton (neutron) rescatters off a neutron (proton) and the latter one is detected. Because of $np$ dominance in the high-$p_{\rm miss}$ regime, SCX has the net effect of decreasing (increasing) the $^3$H ($^3$He) cross section. Further theoretical input is needed to quantify this effect in the kinematics of this measurement.

The E12-14-011 experiment also carried out a search for a bound di-neutron state by comparing $(e,e'p)$ $^3$H and $^3$He distributions to look for signature of a 2-body breakup in $^3$H~\cite{Nguyen:2021spq}. The existence of such state would have profound consequences in nuclear physics~\cite{Li:2022fhh,Hammer:2014rba,Kneller:2003ka}. This analysis placed limits that constrain the $nn$ content of the tritium spectral function to be $<1.5\%$.

\subsection{Inclusive scattering at $x>1$ and short-range correlations}\label{sec:SRC}
The E12-11-112~\cite{E12-11-112} experiment was designed as a precision measurement of the isospin dependence of NN SRCs in A=3 nuclei via $(e,e')$ quasi-elastic (QE) scattering. While for DIS scattering, $x$ represents the longitudinal momentum fraction of the struck quark, for QE scattering it is the kinematic limit of $x$ that is more significant. For scattering from the free proton, $x=1$ is the kinematic limit and elastic scattering corresponds to $x=1$. QE scattering at $x > 1$ requires a non-zero initial nucleon momentum, with scattering up to $x=2$ possible (in principle) for the deuteron, and up to $x \approx A$ for heavier nuclei. By going to $x$ well above unity, one requires a significant initial nucleon momentum. At momenta above the Fermi momentum, these high-momentum nucleons come almost exclusive from short-range correlations~\cite{Arrington:2022rev}, pairs of nucleons that interact via the strong, short-range part of the NN interaction, giving a pair of nucleons with large relative momentum but small total momentum. 

The primary E12-11-112 data involved inclusive cross section measurements from $^2$H, $^3$H, and $^3$He, at $x>1$. These data were taken using a $4.323$ GeV electron beam, with scattered electrons detected by the left HRS at 17.00 or 22.08 degrees. These two data sets covered $x=0.8-3.0$ with $Q^2\approx1.4$~GeV$^2$ in the SRC regime, and $x=1.0-1.8$ with $Q^2\approx1.9$~GeV$^2$ in the SRC regime. The latter data came from the single-arm trigger from the E12-14-011 experiment. 

Based on previous measurements, we know that we can isolate scattering from $k>k_{F}$ in inclusive scattering at $x>$1.4-1.5 for $Q^2>1.4$~GeV$^2$~\cite{Frankfurt:1993sp,Sargsian:2002wc,Fomin:2011ng,Arrington:2022rev}. The cross sections have a universal $x$ and $Q^2$ dependence in these kinematics, based on the underlying two-body structure of the two-nucleon SRCs, and the ratios of these cross sections have been used to map out the relative contribution of SRCs in various nuclei~\cite{Arrington:2011xs,Arrington:2022rev}.  In addition, two-nucleon knockout measurements, where both the struck proton from an SRC and the spectator nucleon are observed~\cite{Shneor:2007tu,Subedi:2008zz,LabHallA:2014wqo,CLAS:2018xvc} have shown that $np$-SRCs strongly dominate over $pp$- and $nn$-SRC for a range of nuclei from $^4$He to $^{208}$Pb~\cite{Hen:2016kwk,Arrington:2022rev}, with $np$-SRCs typically having 30-40 times more contribution than $pp$-SRCs.

\begin{figure}[hbt!]
\centering
\includegraphics[width=0.6\textwidth,height=7cm]{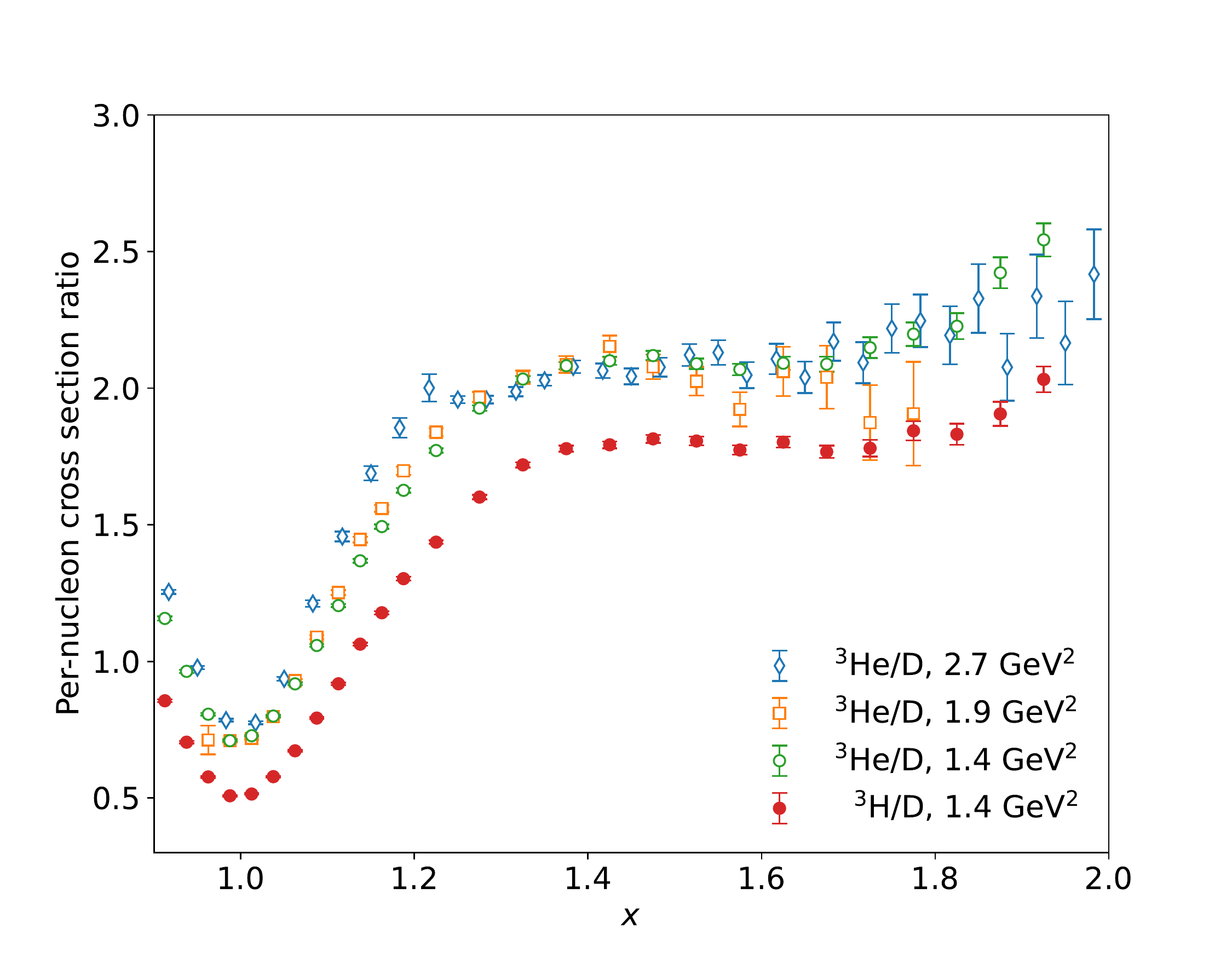}
\includegraphics[width=0.6\textwidth,height=7cm]{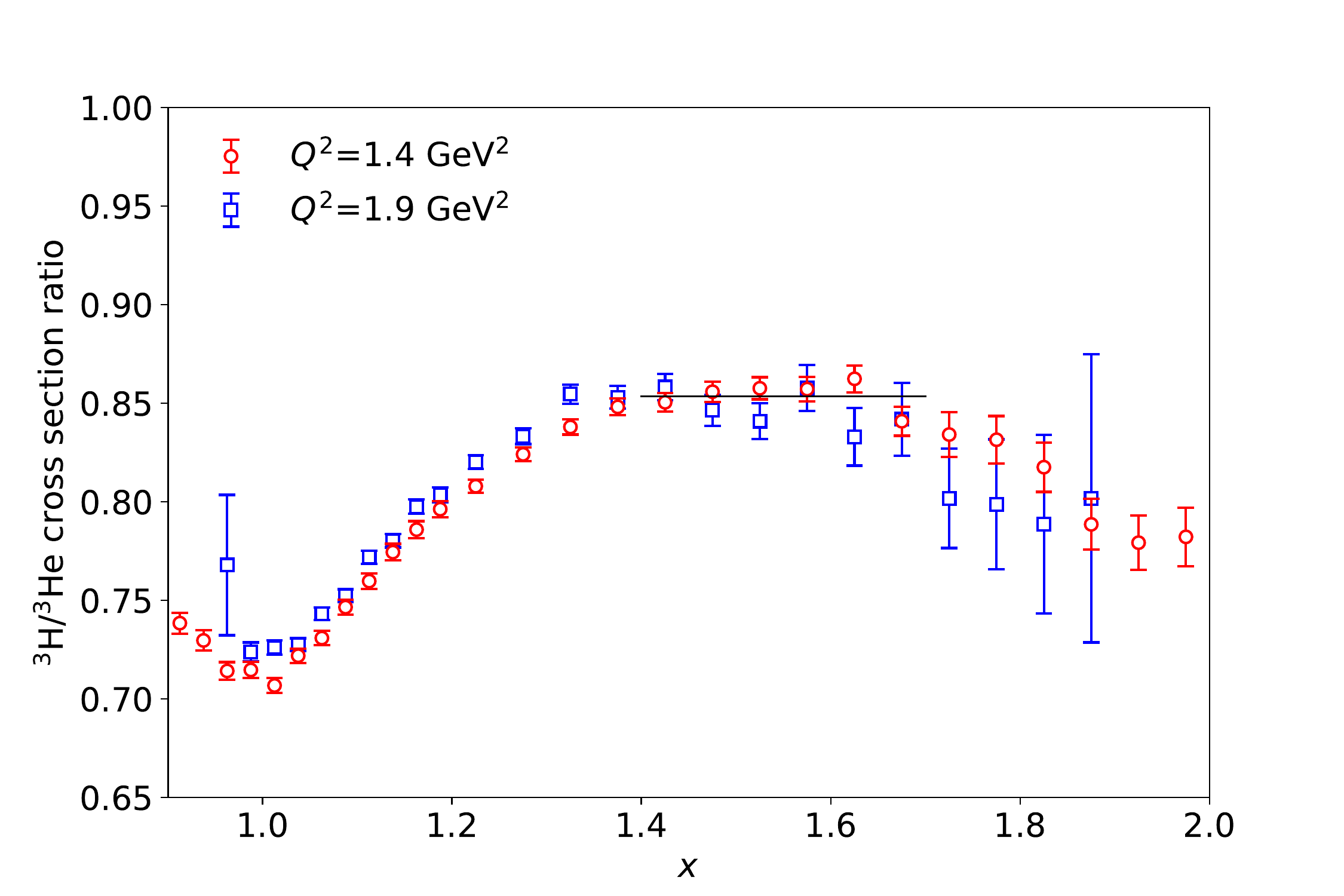}
\caption{[Top] $(e,e')$ per-nucleon cross section ratios of $^3$He or $^3$H to D from JLab experiment E12-11-112 ($Q^2=$$1.4$ GeV$^2$), E12-14-011 ($Q^2=$$1.9$ GeV$^2$) and E02-019 ($Q^2=$$2.7$ GeV$^2$). [Bottom] $(e,e')$ cross section ratios of $^3$H to $^3$He from JLab experiment E12-11-112 ($1.4$ GeV$^2$) and E12-14-011 ($1.9$ GeV$^2$). The horizontal line in black indicates a fit of the plateau between $1.4<x<1.7$ which gives the plateau height of $0.854\pm0.010$. Both figures adapted from Ref.~\cite{Li:2022fhh}. }
\label{fig:tritium_src_ratios}
\small
\end{figure}

While inclusive scattering is sensitive to the overall contribution of SRCs, it is generally not sensitive to the isospin structure of the 2N-SRCs, as there is no detection of the correlated nucleons in the final state. But in a comparison of the mirror nuclei $^3$H and $^3$He, the differing isospin structure of those two targets provides sensitivity to the isospin structure even in inclusive scattering. Following previous inclusive measurements, we take the cross section ratio of a heavier nucleus to the deuteron, and use the plateau region at $x>1.4$ to determine the relative contribution of scattering from the SRCs. Figure~\ref{fig:tritium_src_ratios} shows the $^3$He/$^2$H ratio from the 2018 tritium experiments~\cite{Li:2022fhh} and from the previous JLab Hall C experiment~\cite{Fomin:2011ng}. Both the plateau in $x$ and the $Q^2$ independence of the ratio is observed, as expected in the SRC model~\cite{Frankfurt:1988nt,Frankfurt:1993sp}.

The top panel of Figure~\ref{fig:tritium_src_ratios} shows both the $^3$H/$^2$H and $^3$He/$^2$H ratios from the 1.4~GeV$^2$ measurement of Ref.~\cite{Li:2022fhh}, along with the isoscalar average of the two. The latter provides a measure of the relative number of SRCs in the A=3 nuclei relative to the deuteron, independent of the relative $np$ and $pp$ ($np$ and $nn$ in $^3$H) contributions, with a small correction needed to account for the center-of-mass motion of the SRCs within the nucleus~\cite{Fomin:2011ng}.

\begin{figure}[hbt!]
\centering
\includegraphics[width=0.6\textwidth]{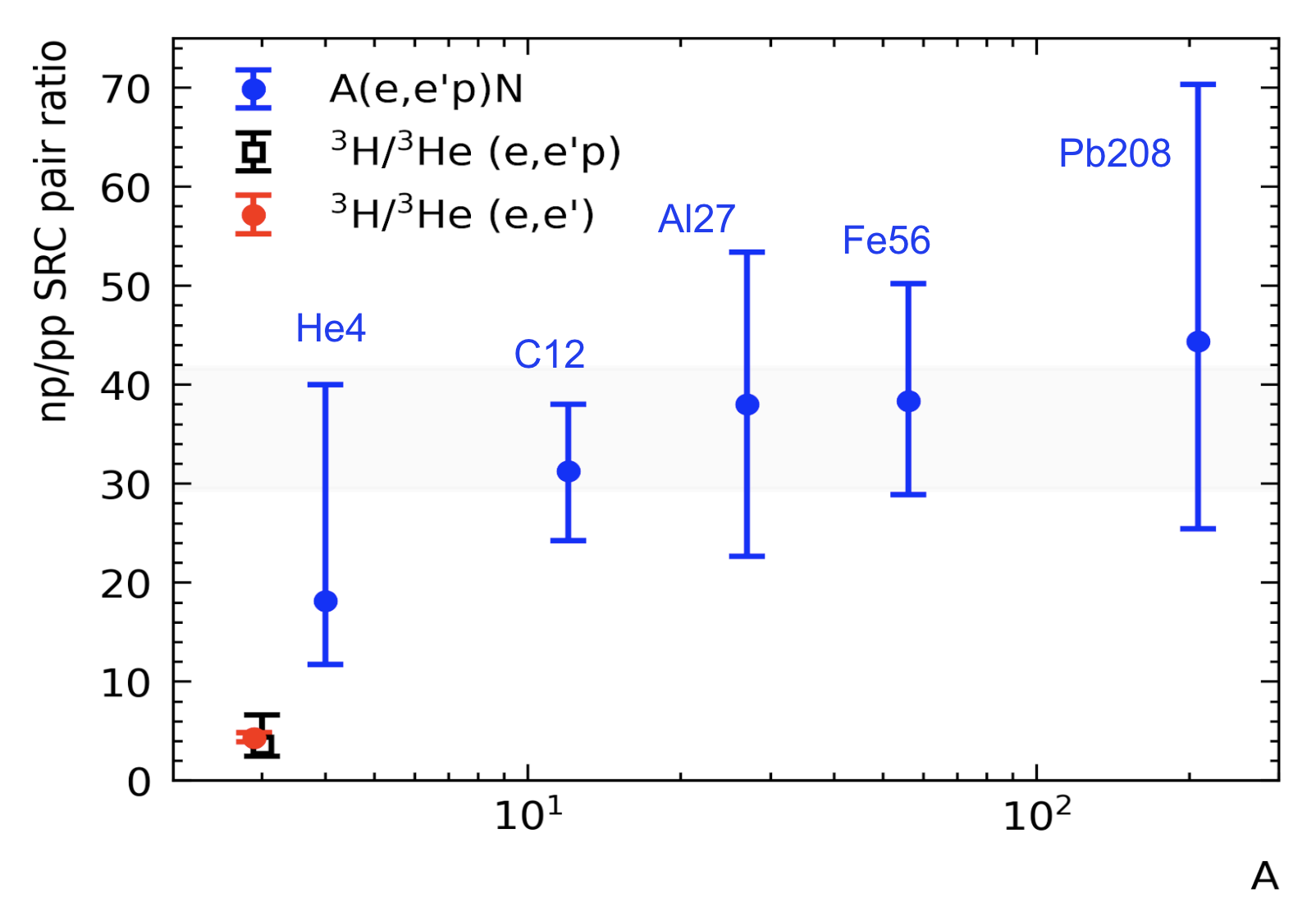}
\caption{ $(e,e')$ $np$-to-$pp$ SRC pair ratios from $A=3$ cross section ratios comparing with extractions from various $(e,e'pN)$ measurements~\cite{Shneor:2007tu,Korover:2014dma,Subedi:2008zz,CLAS:2018xvc}. Figure adapted from Ref.~\cite{Li:2022fhh}.}
\label{fig:tritium_src_final}
\small
\end{figure}

If only $np$-SRCs contribute, which is a good approximation based on the previous measurements in heavier nuclei~\cite{Arrington:2022rev}, the $^3$H/$^2$H and $^3$He/$^2$H cross section ratios would be expected to be identical, as only the $np$-SRCs would contribute, and these would be identical for $^3$H and $^3$He based on isospin symmetry. The fact that the $^3$He/$^2$H ratio is larger suggests that the $pp$-SRCs in $^3$He and $nn$-SRCs in $^3$H have a non-negligible contribution, with the fact that the e-p cross section is a factor of 2.5-3 larger than the e-n cross section, enhancing the inclusive results on $^3$He relative to $^3$H.

Figure~\ref{fig:tritium_src_final} shows the $^3$H/$^3$He ratio for both $Q^2$ values, which yield a consistent ratio of $0.854\pm0.010$ in the 2N-SRC region~\cite{Li:2022fhh}, with deviations from scaling for $x>1.7$ due to the differing contributions of two-body and three-body breakup in $^3$He relative to $^3$H~\cite{Li:2022fhh}. Assuming the probability to form $np$ SRC pairs, $N_{np}$, is the same in A=3 nuclei, and that $N_{pp}^{^3\text{He}}=N_{nn}^{^3\text{H}}$ due to isospin symmetry, we have
\begin{align}
\label{eq:ep_ratio}\frac{\sigma_{^3{\rm H}(e,e'p)}}{\sigma_{^3{\rm He}(e,e'p)}}&=\frac{N_{np}\sigma_{np}}{N_{np}\sigma_{np}+N_{pp}\sigma_{pp}},\\
\label{eq:ee_ratio}\frac{\sigma_{^3{\rm H}(e,e')}}{\sigma_{^3{\rm He}(e,e')}}&=\frac{N_{np}\sigma_{np}+N_{nn}\sigma_{nn}}{N_{np}\sigma_{np}+N_{pp}\sigma_{pp}}.
\end{align}
Assuming the the scattering from the SRC is proportional to the sum of the e-N elastic cross section for each of the nucleons~\cite{Ye:2018jth} (and applying DeForest offshell prescription~\cite{DeForest:1983ahx}), we can go from the measured cross section ratio to a ratio of the number of $np$-SRC to $pp$-SRC pairs, $R_{np/pp}=N_{np}/N_{pp}$. After corrections on the center-of-mass differences between $pp$ ($np$) in $^3$He and $nn$ ($np$) in $^3$H~\cite{Li:2022fhh}, one obtains:
\begin{align}
R_{np/pp}(e,e')  &= 4.34^{+.49}_{-.40},\\
R_{np/pp}(e,e'p) &= 3.57^{+3.10}_{-1.13}   
\end{align}
where for the (e,e'p) extraction, the cross section ratios from Fig.~\ref{fig:eep1} are used to extract $R_{np/pp}$ from Eq.~\ref{eq:ep_ratio}.  These data provide an extremely high-precision extraction of the $np$/$pp$ ratio, and yield significantly smaller ratio than observed in previous measurements, as show in the right panel of Fig.~\ref{fig:tritium_src_final}.

Theoretically, the $x>2$ region allows access to extremely high-momentum nucleons which could be associated with the contribution of a possible three nucleon high-momentum configuration (3N-SRC)~\cite{Frankfurt:1988nt,Frankfurt:1993sp,Arrington:2022rev}. Recent works~\cite{Fomin:2017ydn,PhysRevC.107.014319} suggest that a $Q^2$ of 5 GeV$^2$ or higher may be needed to cleanly isolate 3N SRC in inclusive scattering. While the current $^3$H/$^3$He data sets extend to $x\approx 3$, they are at significantly lower $Q^2$ values. 
 However, the terms that yield scaling violations in A/$^2$H and A/$^3$He ratios have significant cancellation in the $^3$H/$^3$He ratio. The E12-11-112 data include several measurements at lower $Q^2$ values, down to $Q^2\approx 0.6$~GeV$^2$, allowing for us to study these scaling violations in much more detail.

If the scaling violations are shown to be sufficiently small, one may still be able to make conclusions about the structure of 3N-SRCs, although this will yield a model-dependent measurement on 3N-SRC structure. But in the comparison of A=3 systems, one can still extract the momentum and isospin structure of the highest-momentum nucleons, even if one cannot isolate 3N-SRC contributions. While this will be sensitive to a mix of 2N- and 3N-SRC contributions, it will provide useful insight into the isospin structure of the highest-momentum part of the nuclear momentum distribution.

\subsection{Extraction of the neutron magnetic form factor from QE scattering}\label{sec:QE}

In addition to measuring quasi-elastic (QE) scattering from extremely high momentum nucleons, E12-11-112~\cite{E12-11-112} also measured inclusive scattering centered on the QE peak. By integrating over the QE peak, the $^3$H and $^3$He cross sections should be proportional to the sum the e-N elastic scattering cross sections from the two constituent nucleons, and their ratio should be:
\begin{equation}
R = \sigma_{3H}/\sigma_{3He} \approx R_{Free} = (\sigma_{ep}+2\sigma_{en})/(2\sigma_{ep}+\sigma_{en}),
\label{eq:qeratio}
\end{equation}
which depend only on the ratio $\sigma_{en}/\sigma_{ep}$.  By measuring the $^3$H/$^3$He cross section ratio and correcting for the nuclear effects, we can extract the neutron-to-proton elastic cross section ratio, from which we can extract the neutron magnetic form factor.  Data were collected at ten different $Q^2$ settings, from $0.6-2.9$~GeV$^2$. Each of the settings includes two or three HRS central momentum settings to provide full coverage of the central part of the QE peak. The QE data analysis was very similar to inclusive SRC analysis, with luminosity-normalized yields extracted and corrected for the different efficiencies of the detectors and DAQ, livetime, radiative corrections, and backgrounds. The background coming from the walls of the target cell was subtracted by using dummy runs and the tritium data was also corrected for $^3$He contamination.

Data were taken with beam energies of 2.222 and 4.323~GeV, with higher-energy data ($Q^2=1.36$~GeV and $Q^2>2$~GeV)  taken with the second target cell. For this experiment, it was necessary to subtract the $^1$H contribution to the QE peak, as well as accounting for the loss of tritium associated with the hydrogen (water) contamination. The hydrogen contamination was subtracted using measurements taken on an identical hydrogen cell, or simulated elastic cross section yields where data was not available. 

\begin{figure}[!htb]
    \centering
    \includegraphics[width=0.5\textwidth,trim={0 0 0 0},clip]{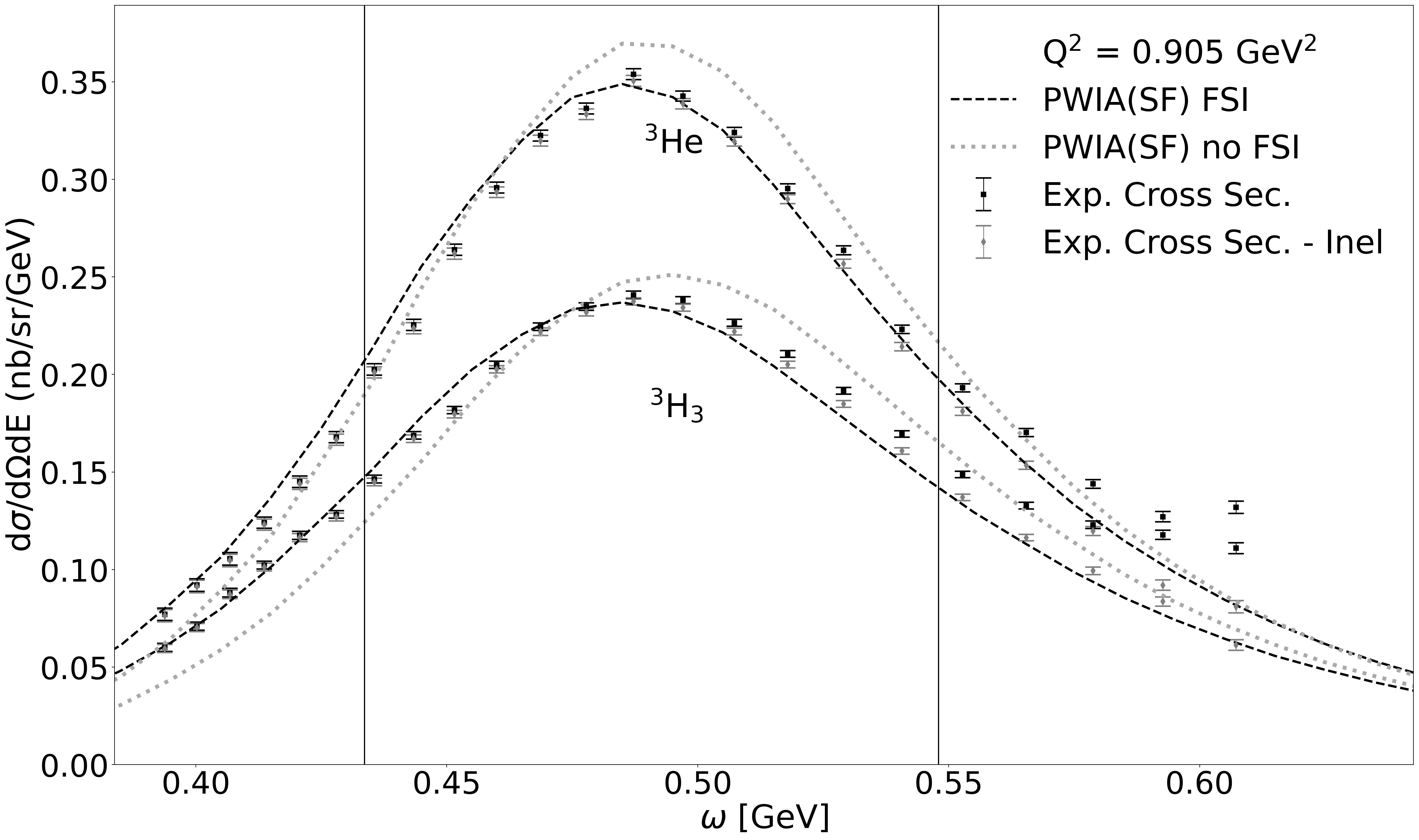}
    \caption{Cross sections and statistical uncertainties for $^3$H and $^3$He compared to calculations based on Ref.~\cite{Andreoli:2021cxo} for $Q^2$ = 0.905 GeV$^2$. The black (grey) points are the cross section before (after) subtraction of the inelastic contribution. The vertical lines represent the integration region to estimate the $^3$H/$^3$He ratio.}
    \label{fig:xs_L28}
\end{figure}

Inelastic background smeared into the QE peak and its contribution increased with $Q^2$. The inelastic contribution was subtracted by using the Bosted-Christy empirical fit~\cite{Christy:2007ve}, with a modified version of the the meson-exchange (ME) corrections. It was shown that the ME contribution was overestimated for light nuclei at small $Q^2$ kinematics, therefore a range of ME calculations~\cite{PhysRevC.104.025501} and parameterizations were used with an additional uncertainty included to account for the model dependence associated with this correction. Fig.\ref{fig:xs_L28} illustrates the QE cross sections for $Q^2$ =  0.905 GeV$^2$ with and without subtraction of the inelastic contribution. It also shows the calculations based on Ref.~\cite{Andreoli:2021cxo} which will be used to estimate the medium effects, and the vertical lines represent the integration region. 

After isolating the QE contribution, the cross section was integrated over $\pm 1 \sigma$ region. This is corrected for nuclear effects using the \textit{ab initio}-based calculations of Ref.~\cite{Andreoli:2021cxo}. From this, we obtain the neutron-to-proton cross section ratio from Eq.~\ref{eq:qeratio}, which is converted to the neutron cross section by using the elastic proton cross section from Ref.~\cite{Arrington:2007ux}, table IV, which does not apply two-photon exchange (TPE) corrections. After applying TPE corrections to the neutron cross section to obtain the Born cross section~\cite{arrington:2011dn}, the magnetic form factor is extracted by subtracting the small contribution from the charge form factor, 
where $G_{E}^{n}$ and its uncertainties are taken from the parametrization of Ref.~\cite{YE20188}. 

\begin{figure}[!htb]
    \centering
    \includegraphics[width=0.53\textwidth]{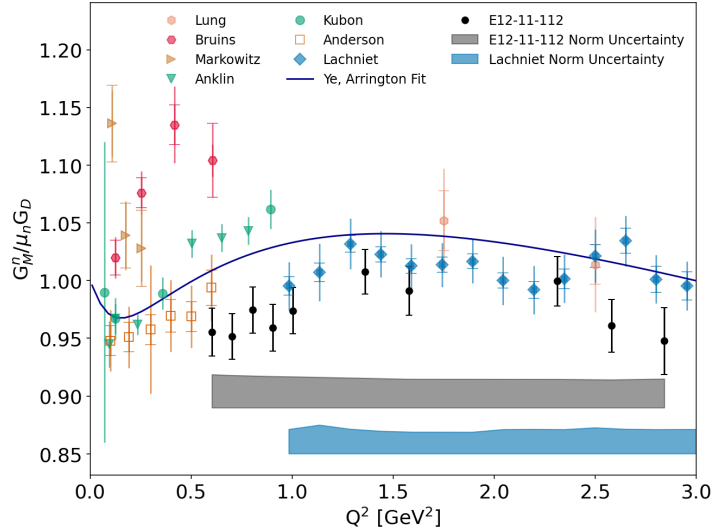}
    \caption{The black points show the preliminary $G_M^n$ results, \textbf{extracted without nuclear corrections}, but with estimates of both the systematic and model-dependence uncertainties. The shaded bands indicate correlated uncertainties. The other points correspond to measurement from previous experiments~\cite{PhysRevLett.70.718,PhysRevLett.75.21,PhysRevC.48.R5,ANKLIN1994313,ANKLIN1998248,KUBON200226,PhysRevC.75.034003,PhysRevLett.102.192001}. The solid line corresponds to a fit to world's electron scattering data~\cite{YE20188}. }
    \label{fig:GMN_datrat}
\end{figure}

Fig.~\ref{fig:GMN_datrat} shows the \textbf{preliminary} E12-11-112 results with no correction applied for the different nuclear effects in $^3$H and $^3$He. Medium corrections are expected to raise these points by several percent, and an estimate of the model-dependence has been included in the uncertainties shown.

\subsection{Nuclear elastic scattering and the triton charge radius}\label{sec:elastic}

The final tritium measurement related to conventional nucleon and nuclear structure was a low-energy measurement of nuclear elastic scattering focused on the extraction of the nuclear charge radius.  In particular, the difference between the charge radius in $^3$H and $^3$He. This corresponds closely to the difference between the radius of proton distributions in the nuclei, and by isospin symmetry, the difference between the proton and neutron distributions in the A=3 system.

In the non-relativistic limit, the form factors are the Fourier transforms of the spatial charge and magnetization distributions inside the nucleus. In this case, the form factors can be expanded in terms of the moments of the charge (magnetization) distributions, e.g.,
\begin{equation}
  F(q^2)\approx 1- \frac{q^2}{6\hbar^2} \langle r^2\rangle +   \frac{q^4}{5!~\hbar^2} \langle r^4\rangle ....
  \label{eq:formfactorf}
\end{equation}
By measuring the electric form factor of the nuclear target at low $q^2$, the square of the transferred three-momentum, one can extract the nuclear charge radius from the slope of the electric form factor,
 \begin{equation}
\langle r^2\rangle \approx -{6\hbar^2} \left.\frac{dF_{ch}(q^2)}{dq^2}\right|_{q^2=0}
\end{equation}
While this is derived in the non-relativistic limit, this is conventionally taken as the definition the RMS charge radius in the full relativistic case. 

Experiment E12-10-003~\cite{E12-10-003} proposed to make such a measurement at a very low $Q^2$ value, where the falloff of the form factor was very sensitive to the nuclear charge radius. In this case, the form factor ratio can be directly converted to a difference in the charge radii, with modest uncertainties associated with the assumptions made about the underlying shape of the charge distributions. While this experiment was not run during the 2018 tritium run group, some data at low energy was taken, corresponding to larger $Q^2$ values where the form factor is not as directly sensitive to the radius. 

At the larger $Q^2$ values measured, the individual $^3$H and $^3$He form factor measurement provide only limited improvement in our extraction of the charge radii of the nuclei. For $^3$He, there is much more data, and more precise measurements, and adding this one additional point adds very little. But because existing $^3$H data are much more limited, the charge radius extraction is sensitive to normalization uncertainties of the data sets, which can be large. Thus, even a single $^3$H point can improve the radius extraction if it's uncertainties are small enough to help constrain the normalization offsets of previous measurements. However, it is possible to do better by using the precise \textit{ratio} of the $^3$H/$^3$He cross section to constrain the $^3$H radius.  The ratio has much smaller uncertainties than the individual cross sections, due to the cancellation of most of the dominant systematic uncertainties. This can be combined with the precise $^3$He form factor measurements to yield an extraction of the $^3$H form factor with smaller uncertainties than the direct cross section extraction. The ratio could also be used as a direct input to a \textit{simultaneous} fit of the $^3$H and $^3$He form factors, allowing for an improved extraction of the difference in their charge radii.

In the Born approximation, the elastic scattering cross section for these spin-1/2 nuclei depends on the Mott cross section and the electric ($F_{ch}$) and magnetic ($F_m$) form factors of nuclei,
\begin{equation}
  \left(\frac{d\sigma}{d\Omega}\right)= \left(\frac{d\sigma}{d\Omega}\right)_{Mott}\frac{1}{\eta} \bigg[\frac{Q^2}{q^2}F^2_{ch}(q)+\frac{\mu^2 Q^2}{2M^2}\left(\frac{1}{2}\frac{Q^2}{q^2}+ \tan^2 \left(\frac{\theta}2\right)\right)F^2_{m}(q)\bigg]
  \label{eq:born cross-section}
\end{equation}
where $\eta= 1+ \frac{Q^2}{4M^2}$ and $\mu$ is the magnetic moment of the target.

We collected data at a beam energy of 1.171~GeV and scattering angle of~$17^{\circ}$, corresponding to $Q^2 = 0.11$~GeV$^2$. The scattering electrons from $^3$H and  $^3$He targets were measured to extract the normalized yield of elastic scattering events.  This was compared to a simulation of the measurement, accounting for acceptance, radiative corrections, and energy loss, to extract the underlying cross section. For both targets, the elastic cross section is extracted with an uncertainty of roughly 4.5\%, with the acceptance, integrated charge measurement, and radiative corrections yielding the dominant sources of uncertainty. The uncertainty in the integrated charge measurement is larger than for the other measurement on tritium, as the beam current was low, typically $\sim$5~$\mu$A, to keep the rate in the spectrometer reasonable. This makes extracting the beam charge much more sensitive to the details of the calibration procedure which is optimized for larger beam currents. However, when we take the cross section ratios - discussed later - all of the large contributions to the uncertainty partially or completely cancel, providing a more precise measurement of the cross section ratio. 

As seen in Eq.~\ref{eq:born cross-section}, both the charge and magnetic form factors contribute to the cross section. In order to isolate the charge form factor, the magnetic contribution needs to be removed. We use the fit to the magnetic form factor from Ref.~\cite{AMROUN1994596} to obtain $F_m$ and calculate its contribution to the elastic cross section. We subtract this to isolate the charge contribution, applying a 5\% uncertainty to the value of $F_m$ used in the subtraction, roughly twice the uncertainty quoted in~\cite{AMROUN1994596}, to be conservative. From this we can extract $F_{ch}$ for both $^3$H and $^3$He with a precision of around 2.5\%.

In taking the cross section ratios, the uncertainties associated with most of the corrections that are applied have significant cancellation, and the final $^3$H/$^3$He cross section ratio is extracted with an uncertainty that is statistics dominated and less than 2\%. The systematic uncertainty is dominated by the 1.06\% uncertainty on the relative thickness of the two targets, with the charge measurement, radiative corrections, and relative acceptance all making contributions at or below the 0.3\% level. After correcting for the magnetic contributions, we extract the ratio of the electric contributions to the cross section to better than 3\%, yielding a measurement of $F_{ch}$($^3$H)/$F_{ch}$($^3$He) to better than 1.5\%.  

\begin{figure}[hbt!]
\centering
\includegraphics[width=0.51\textwidth]{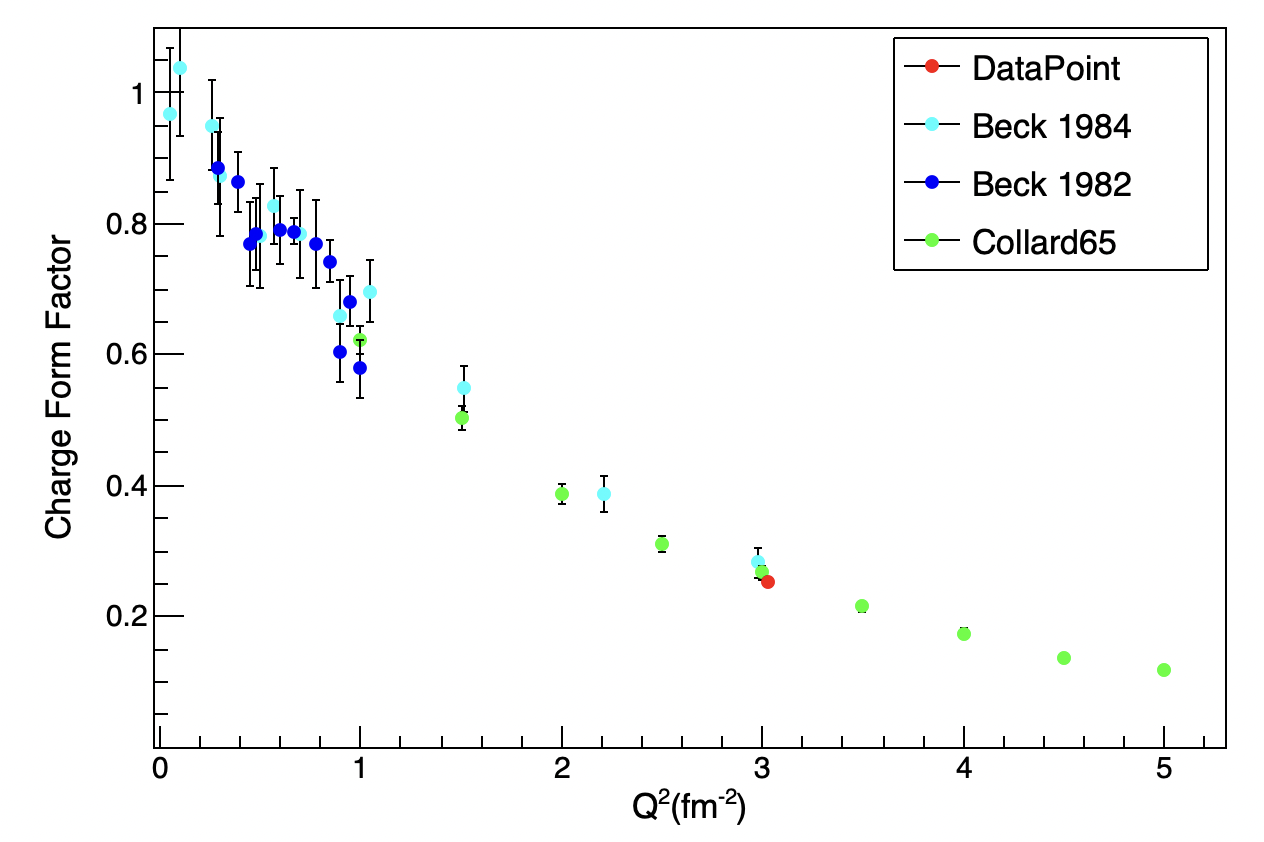}
\includegraphics[width=0.48\textwidth]{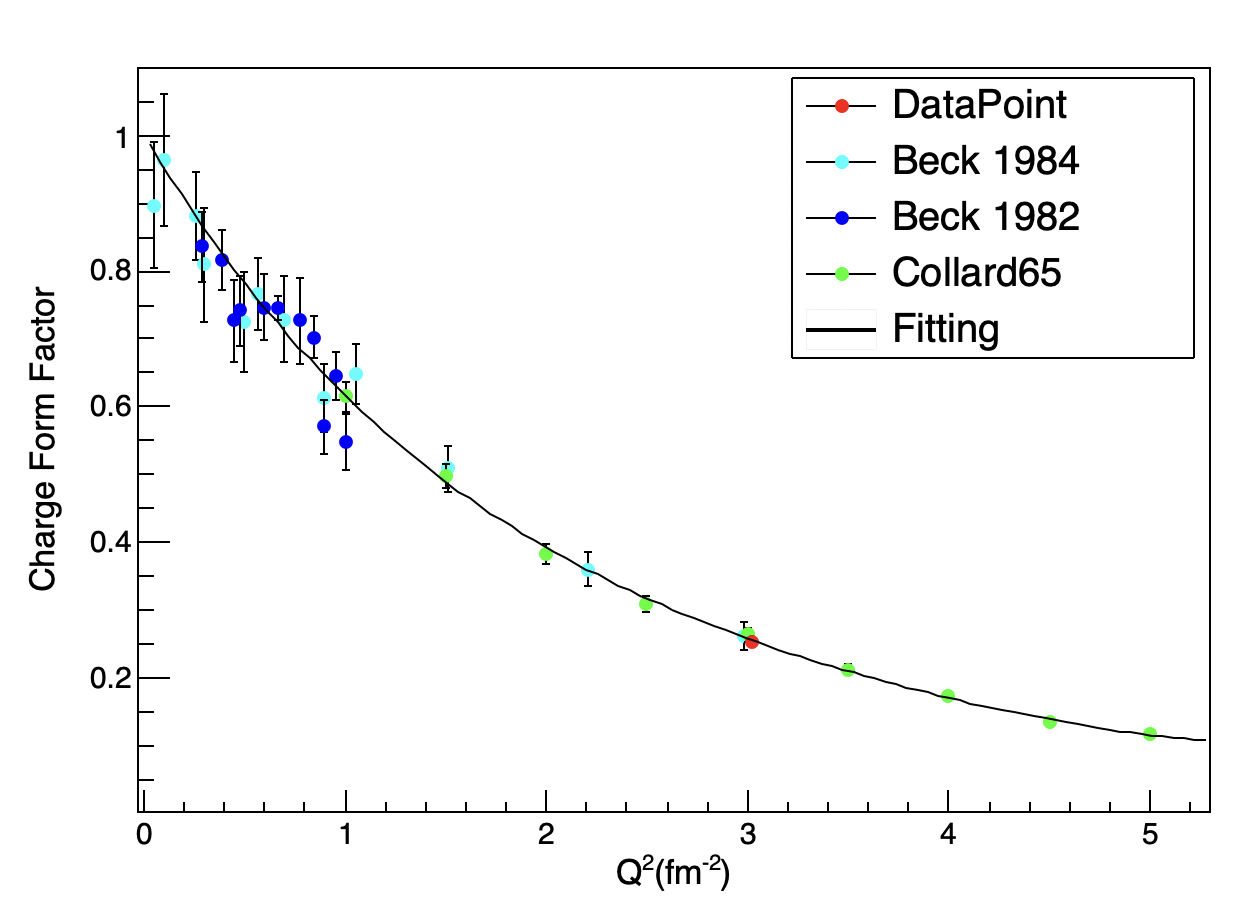}
\caption{[Left] F$_{ch}$($^3$H) world data sets, along with the current preliminary results. [Right] The same data, along with a fit (4-degree polynomial function) to the data. The experimental normalization factors are allowed to vary as part of the fit, and the data are shown with the normalization factors obtained in the fit applied.}
\label{fitting1}
\small
\end{figure}

There are many studies for $F_{ch}$($^3$He) and their results showed a good agreement with each other and with the theoretical studies. The most complete analysis was done by Amroun et al.~\cite{AMROUN1994596}, who collected many world data sets and performed a global fit, which provides $F_{ch}$($^3$He) at $Q^2=0.11$~GeV$^2$ with an 0.5\% uncertainty. Our extracted $F_{ch}$($^3$H)/$F_{ch}$($^3$He) ratio was multiplied by $F_{ch}$($^3$He) from~\cite{AMROUN1994596} to obtain $F_{ch}$($^3$H), shown in Figure~\ref{fitting1}, with fractional uncertainty of $\sim$1.4\%. 

To evaluate the impact of the new measurement, we perform a simple fit to the $^3$H form factor data at low $Q^2$, using experiments which quote results for $F_{ch}$~\cite{Beck:1984qhv,6662e8b2e0a649be97d95883e32b1ccd,PhysRev.138.B57}. The data, shown in Fig.~\ref{fitting1}, are fit with a 4th order polynomial, allowing for the previous data sets to vary within their estimated normalization uncertainties.  This fit is performed with and without our data point at $Q^2=0.11$~GeV$^2$, and we estimate the impact of the new data point by comparing the final uncertainties obtained for the normalization factors of the previous data sets and the slope of the fit at $Q^2=0$. Including our new data point yields small changes in the normalization factors of the previous data sets, and decreases the associated uncertainty of these normalization factors by $\sim$30-50\%.  It also reduces the uncertainty on the parameters of the polynomial fit by $\sim$30\%. This is not intended to be a full analysis or extraction of the radii; it simply illustrates the way in which this measurement, even at $Q^2$ values far from the linear falloff region, can improve our understanding of the charge radius.

\section{Summary}

In conclusion, the new tritium target designed for use at Jefferson Lab led to an extremely successful program focused on a detailed understanding of the structure, and in particular isospin structure, of A=3 nuclei. Measurements took advantage of the isospin structure of the mirror nuclei to make precision comparisons of the neutron-to-proton cross sections, allowing for improved extractions of the neutron structure function and the neutron magnetic form factor, as well as providing a new comparison of the nuclear elastic form factors for $^3$H and $^3$He.  In addition, inclusive and single-nucleon knockout measurements from $^3$H and $^3$He provided insight into the proton and neutron momentum distributions at low momenta, while providing constraints for both the neutron-to-proton ratio at high momentum and final-state interactions for (e,e'p) measurements at large missing momentum.  The studies of the high-momentum structure led to completely new information on the isospin structure of short-range correlations in A=3 nuclei, which had never been explored before. While the program focused on nucleon and nuclear structure, a measurement involving kaon detection provided the opportunity to look for production of $\Lambda$ hypernuclei in scattering from $^3$H, allowing tests of the $\Lambda$-N interaction.  There have already been multiple proposals and letters of intent that aim to bring back these targets to further extend such measurements. 

\begin{acknowledgements}
The tritium program would not have been possible without the work of the tritium target task force, including Roy Holt and Patricia Solvignon, the Target Group at Jefferson Lab, and the support from Savannah River Tritium Enterprises. We wish to thank the Jefferson Lab Hall A scientific and engineering staff for their outstanding support, especially Douglas Higinbotham for coordinating all tritium experiments. This work was supported by U.S. Department of Energy, Office of Science, Office of Nuclear Physics, under contracts DE-AC02-05CH11231 and DE-FG02-88ER40410, and contract DE-AC05-06OR23177 under which JSA, LLC operates Jefferson Lab.

\end{acknowledgements}

\bibliographystyle{spphys}       
\bibliography{Tritium_EPJA}   

\begin{thebibliography}{100}
\providecommand{\url}[1]{{#1}}
\providecommand{\urlprefix}{URL }
\expandafter\ifx\csname urlstyle\endcsname\relax
  \providecommand{\doi}[1]{DOI \discretionary{}{}{}#1}\else
  \providecommand{\doi}{DOI \discretionary{}{}{}\begingroup
  \urlstyle{rm}\Url}\fi

\bibitem{Arrington:2011qt}
J.~Arrington, J.G. Rubin, W.~Melnitchouk, Phys. Rev. Lett. \textbf{108}, 252001
  (2012)

\bibitem{Accardi:2011fa}
A.~Accardi, W.~Melnitchouk, J.F. Owens, M.E. Christy, C.E. Keppel, L.~Zhu, J.G.
  Morfin, Phys. Rev. D \textbf{84}, 014008 (2011)

\bibitem{Arrington:2006zm}
J.~Arrington, C.D. Roberts, J.M. Zanotti, J. Phys. G \textbf{34}, S23 (2007)

\bibitem{Afnan:2003vh}
I.R. Afnan, et~al., Phys. Rev. C \textbf{68}, 035201 (2003)

\bibitem{Wiringa:2013ala}
R.B. Wiringa, R.~Schiavilla, S.C. Pieper, J.~Carlson, Phys. Rev. C \textbf{89},
  024305 (2014)

\bibitem{Arrington:2022rev}
J.~Arrington, N.~Fomin, A.~Schmidt, Annual Review of Nuclear and Particle
  Science \textbf{72}, 307 (2022)

\bibitem{E12-17-003}
L.~Tang, et~al.
\newblock {E12-17-003: Determining the Unknown $\Lambda{}-n$ Interaction by
  Investigating the $\Lambda{}nn$ Resonance} (2017).
\newblock
  \urlprefix\url{https://www.jlab.org/exp\_prog/proposals/17/PR12-17-003.pdf}

\bibitem{HallA:2022qqj}
B.~Pandey, et~al., Phys. Rev. C \textbf{105}, 051001 (2022)

\bibitem{Suzuki:2021pid}
K.N. Suzuki, et~al., PTEP \textbf{2022}, 013D01 (2022)

\bibitem{Hughes:1966zza}
E.B. Hughes, M.R. Yearian, R.~Hofstadter, Phys. Rev. \textbf{151}, 841 (1966)

\bibitem{Beck:1984qhv}
D.H. Beck, et~al., Phys. Rev. C \textbf{30}, 1403 (1984)

\bibitem{Beck:1989bi}
D.~Beck, et~al., Nucl. Instrum. Meth. A \textbf{277}, 323 (1989)

\bibitem{Juster:1985sd}
F.P. Juster, et~al., Phys. Rev. Lett. \textbf{55}, 2261 (1985)

\bibitem{target_NIM}
B.~Brajuskovic, T.~O'Connor, R.~Holt, J.~Reneker, D.~Meekins, P.~Solvignon,
  Nucl. Instrum. Meth. A \textbf{729}, 469 (2013)

\bibitem{Alcorn:2004sb}
J.~Alcorn, et~al., Nucl. Instrum. Meth. A \textbf{522}, 294 (2004)

\bibitem{Santiesteban:2018qwi}
S.N. Santiesteban, et~al., Nucl. Instrum. Meth. A \textbf{940}, 351 (2019)

\bibitem{E12-10-103}
G.~Petratos, J.~Arrington, J.~Gomez, M.~Katramatou, D.~Meekins, R.~Ransome.
\newblock {E12-10-103: Measurement of the F2n/F2p, d/u Ratios and A=3 EMC
  Effect in Deep Inelastic Scattering off the Tritium and Helium Mirror Nuclei}
  (2010).
\newblock
  \urlprefix\url{https://www.jlab.org/exp\_prog/proposals/10/PR12-10-103.pdf}

\bibitem{E12-11-112}
J.~Arrington, D.~Day, D.~Higinbotham, Z.~Ye.
\newblock {E12-11-112: Precision measurement of the isospin dependence in the
  2N and 3N short range correlation region} (2011).
\newblock
  \urlprefix\url{https://www.jlab.org/exp\_prog/proposals/11/PR12-11-112.pdf}

\bibitem{E12-14-011}
{L. Weinstein, W. Boeglin, O. Chen, F. Hauenstein}.
\newblock {E12-14-011: Proton and neutron momentum distributions in A=3
  asymmetric nuclei} (2011).
\newblock
  \urlprefix\url{https://www.jlab.org/exp\_prog/proposals/14/PR12-14-011.pdf}

\bibitem{E12-10-003}
D.~Higinbotham, J.~Arrington, T.~Averett, L.~Myers.
\newblock {E12-10-003: Ratio of the electric form factor in the mirror nuclei
  3He and 3H} (2010).
\newblock
  \urlprefix\url{https://www.jlab.org/exp\_prog/proposals/10/PR12-10-003.pdf}

\bibitem{Achenbach:2023pba}
P.~Achenbach, et~al., arXiv:2303.02579  (2023)

\bibitem{E12-20-005}
H.~Szumilla-Vance, et~al.
\newblock {E12-20-005: Precision measurements of A=3 nuclei in Hall B} (2020).
\newblock
  \urlprefix\url{https://www.jlab.org/exp\_prog/proposals/20/PR12-20-005.pdf}

\bibitem{E12-21-004}
L.~Weinstein, et~al.
\newblock {E12-21-004: Semi-Inclusive Deep Inelastic Scattering Measurement of
  A=3 Nuclei with CLAS12 in Hall B} (2021).
\newblock
  \urlprefix\url{https://www.jlab.org/exp\_prog/proposals/21/PR12-21-004.pdf}

\bibitem{Arrington:2021alx}
J.~Arrington, et~al., Prog. Part. Nucl. Phys. \textbf{127}, 103985 (2022)

\bibitem{Geesaman:1995yd}
D.F. Geesaman, K.~Saito, A.W. Thomas, Ann. Rev. Nucl. Part. Sci. \textbf{45},
  337 (1995).
\newblock \doi{10.1146/annurev.ns.45.120195.002005}

\bibitem{E140X:1995ims}
L.H. Tao, et~al., Z. Phys. C \textbf{70}, 387 (1996).
\newblock \doi{10.1007/s002880050117}

\bibitem{E12-14-002}
S.P. Malace, M.E. Christy, D.~Gaskell, C.~Keppel, P.~Solvignon.
\newblock {E12-14-002: Precision Measurements and Studies of a Possible Nuclear
  Dependence of $R=\sigma_L/\sigma_T$} (2014).
\newblock
  \urlprefix\url{https://www.jlab.org/exp\_prog/proposals/14/PR12-14-002.pdf}

\bibitem{Abrams:2021xum}
D.~Abrams, et~al., Phys. Rev. Lett. \textbf{128}, 132003 (2022)

\bibitem{Kulagin:2004ie}
S.A. Kulagin, R.~Petti, Nucl. Phys. A \textbf{765}, 126 (2006)

\bibitem{Kulagin:2010gd}
S.A. Kulagin, R.~Petti, Phys. Rev. C \textbf{82}, 054614 (2010)

\bibitem{Alekhin:2017fpf}
S.I. Alekhin, S.A. Kulagin, R.~Petti, Phys. Rev. D \textbf{96}, 054005 (2017)

\bibitem{Bodek:1979rx}
A.~Bodek, et~al., Phys. Rev. D \textbf{20}, 1471 (1979)

\bibitem{CLAS:2014jvt}
S.~Tkachenko, et~al., Phys. Rev. C \textbf{89}, 045206 (2014).
\newblock [Addendum: Phys.Rev.C 90, 059901 (2014)]

\bibitem{Valenty:2022gqm}
H.~Valenty, J.R. West, F.~Benmokhtar, D.W. Higinbotham, A.~Parker, E.~Seroka,
  arXiv:2210.04372  (2022)

\bibitem{Ding:2022ows}
M.~Ding, C.D. Roberts, S.M. Schmidt, Particles \textbf{6}, 57 (2023)

\bibitem{Segarra:2021exb}
E.P. Segarra, et~al., arXiv:2104.07130  (2021)

\bibitem{Alekhin:2022tip}
S.I. Alekhin, S.A. Kulagin, R.~Petti, Phys. Rev. D \textbf{105}, 114037 (2022)

\bibitem{Alekhin:2022uwc}
S.I. Alekhin, S.A. Kulagin, R.~Petti, arXiv:2211.09514  (2022)

\bibitem{Cocuzza:2021rfn}
C.~Cocuzza, C.E. Keppel, H.~Liu, W.~Melnitchouk, A.~Metz, N.~Sato, A.W. Thomas,
  Phys. Rev. Lett. \textbf{127}, 242001 (2021)

\bibitem{DeForest:1983ahx}
T.~De~Forest, Nucl. Phys. A \textbf{392}, 232 (1983)

\bibitem{Lapikas:1993uwd}
L.~Lapikas, Nucl. Phys. A \textbf{553}, 297c (1993)

\bibitem{JeffersonLabHallA:2004tej}
F.~Benmokhtar, et~al., Phys. Rev. Lett. \textbf{94}, 082305 (2005)

\bibitem{Sargsian:2001ax}
M.M. Sargsian, Int. J. Mod. Phys. E \textbf{10}, 405 (2001)

\bibitem{Sargsian:2002wc}
M.M. Sargsian, et~al., J. Phys. G \textbf{29}, R1 (2003)

\bibitem{Frankfurt:1996xx}
L.L. Frankfurt, M.M. Sargsian, M.I. Strikman, Phys. Rev. C \textbf{56}, 1124
  (1997)

\bibitem{Laget:2004sm}
J.M. Laget, Phys. Lett. B \textbf{609}, 49 (2005)

\bibitem{Jeschonnek:2008zg}
S.~Jeschonnek, J.W. Van~Orden, Phys. Rev. C \textbf{78}, 014007 (2008)

\bibitem{CLAS:2007tee}
K.S. Egiyan, et~al., Phys. Rev. Lett. \textbf{98}, 262502 (2007)

\bibitem{Sargsian:2009hf}
M.M. Sargsian, Phys. Rev. C \textbf{82}, 014612 (2010)

\bibitem{HallA:2011gjn}
W.U. Boeglin, et~al., Phys. Rev. Lett. \textbf{107}, 262501 (2011)

\bibitem{Arrington:2011xs}
J.~Arrington, D.W. Higinbotham, G.~Rosner, M.~Sargsian, Prog. Part. Nucl. Phys.
  \textbf{67}, 898 (2012)

\bibitem{JeffersonLabHallATritium:2019xlj}
R.~Cruz-Torres, et~al., Phys. Lett. B \textbf{797}, 134890 (2019)

\bibitem{JeffersonLabHallATritium:2020mha}
R.~Cruz-Torres, et~al., Phys. Rev. Lett. \textbf{124}, 212501 (2020)

\bibitem{Lonardoni:2018sqo}
D.~Lonardoni, S.~Gandolfi, X.B. Wang, J.~Carlson, Phys. Rev. C \textbf{98},
  014322 (2018)

\bibitem{Wiringa:1994wb}
R.B. Wiringa, V.G.J. Stoks, R.~Schiavilla, Phys. Rev. C \textbf{51}, 38 (1995)

\bibitem{Lonardoni:2018nob}
D.~Lonardoni, S.~Gandolfi, J.E. Lynn, C.~Petrie, J.~Carlson, K.E. Schmidt,
  A.~Schwenk, Phys. Rev. C \textbf{97}, 044318 (2018)

\bibitem{Gezerlis:2014zia}
A.~Gezerlis, et~al., Phys. Rev. C \textbf{90}, 054323 (2014)

\bibitem{Lynn:2015jua}
J.E. Lynn, et~al., Phys. Rev. Lett. \textbf{116}, 062501 (2016)

\bibitem{Lynn:2017fxg}
J.E. Lynn, et~al., Phys. Rev. C \textbf{96}, 054007 (2017)

\bibitem{Lonardoni:2017hgs}
D.~Lonardoni, J.~Carlson, S.~Gandolfi, J.E. Lynn, K.E. Schmidt, A.~Schwenk,
  X.~Wang, Phys. Rev. Lett. \textbf{120}, 122502 (2018)

\bibitem{Kievsky:2008es}
A.~Kievsky, S.~Rosati, M.~Viviani, L.E. Marcucci, L.~Girlanda, J. Phys. G
  \textbf{35}, 063101 (2008)

\bibitem{Marcucci:2018llz}
L.E. Marcucci, F.~Sammarruca, M.~Viviani, R.~Machleidt, Phys. Rev. C
  \textbf{99}, 034003 (2019)

\bibitem{Machleidt:2000ge}
R.~Machleidt, Phys. Rev. C \textbf{63}, 024001 (2001)

\bibitem{Coon:2001pv}
S.A. Coon, H.K. Han, Few Body Syst. \textbf{30}, 131 (2001)

\bibitem{Entem:2017gor}
D.R. Entem, R.~Machleidt, Y.~Nosyk, Phys. Rev. C \textbf{96}, 024004 (2017)

\bibitem{CiofidegliAtti:2004jg}
C.~Ciofi~degli Atti, L.P. Kaptari, Phys. Rev. C \textbf{71}, 024005 (2005)

\bibitem{Benhar:1993ja}
O.~Benhar, V.R. Pandharipande, Phys. Rev. C \textbf{47}, 2218 (1993)

\bibitem{Golak:2005iy}
J.~Golak, et~al., Phys. Rept. \textbf{415}, 89 (2005)

\bibitem{Carasco:2003us}
C.~Carasco, et~al., Phys. Lett. B \textbf{559}, 41 (2003)

\bibitem{code:sargsian_fsi}
M.M. Sargsian.
\newblock private communication

\bibitem{Sargsian:2004tz}
M.M. Sargsian, T.V. Abrahamyan, M.I. Strikman, L.L. Frankfurt, Phys. Rev. C
  \textbf{71}, 044614 (2005)

\bibitem{Sargsian:2005ru}
M.M. Sargsian, T.V. Abrahamyan, M.I. Strikman, L.L. Frankfurt, Phys. Rev. C
  \textbf{71}, 044615 (2005)

\bibitem{Nguyen:2021spq}
D.~Nguyen, et~al., Phys. Lett. B \textbf{831}, 137165 (2022)

\bibitem{Li:2022fhh}
S.~Li, et~al., Nature \textbf{609}(7925), 41 (2022)

\bibitem{Hammer:2014rba}
H.W. Hammer, S.~K\"onig, Phys. Lett. B \textbf{736}, 208 (2014)

\bibitem{Kneller:2003ka}
J.P. Kneller, G.C. McLaughlin, Phys. Rev. D \textbf{70}, 043512 (2004)

\bibitem{Frankfurt:1993sp}
L.L. Frankfurt, M.I. Strikman, D.B. Day, M.~Sargsian, Phys. Rev. C \textbf{48},
  2451 (1993)

\bibitem{Fomin:2011ng}
N.~Fomin, et~al., Phys. Rev. Lett. \textbf{108}, 092502 (2012)

\bibitem{Shneor:2007tu}
R.~Shneor, et~al., Phys. Rev. Lett. \textbf{99}, 072501 (2007)

\bibitem{Subedi:2008zz}
R.~Subedi, et~al., Science \textbf{320}, 1476 (2008)

\bibitem{LabHallA:2014wqo}
I.~Korover, et~al., Phys. Rev. Lett. \textbf{113}, 022501 (2014)

\bibitem{CLAS:2018xvc}
M.~Duer, et~al., Phys. Rev. Lett. \textbf{122}, 172502 (2019)

\bibitem{Hen:2016kwk}
O.~Hen, G.A. Miller, E.~Piasetzky, L.B. Weinstein, Rev. Mod. Phys. \textbf{89},
  045002 (2017)

\bibitem{Frankfurt:1988nt}
L.L. Frankfurt, M.I. Strikman, Phys. Rept. \textbf{160}, 235 (1988)

\bibitem{Korover:2014dma}
I.~Korover, et~al., Phys. Rev. Lett. \textbf{113}, 022501 (2014)

\bibitem{Ye:2018jth}
Z.~Ye, J.~Arrington, arXiv:1810.03667  (2018).
\newblock {Proceedings of the 13th Conference on the Intersections of Particle
  and Nuclear Physics}

\bibitem{Fomin:2017ydn}
N.~Fomin, D.~Higinbotham, M.~Sargsian, P.~Solvignon, Ann. Rev. Nucl. Part. Sci.
  \textbf{67}, 129 (2017)

\bibitem{PhysRevC.107.014319}
D.B. Day, L.L. Frankfurt, M.M. Sargsian, M.I. Strikman, Phys. Rev. C
  \textbf{107}, 014319 (2023)

\bibitem{Christy:2007ve}
M.E. Christy, P.E. Bosted, Phys. Rev. C \textbf{81}, 055213 (2010)

\bibitem{PhysRevC.104.025501}
V.L. Martinez-Consentino, I.~Ruiz~Simo, J.E. Amaro, Phys. Rev. C \textbf{104},
  025501 (2021)

\bibitem{Andreoli:2021cxo}
L.~Andreoli, J.~Carlson, A.~Lovato, S.~Pastore, N.~Rocco, R.B. Wiringa, Phys.
  Rev. C \textbf{105}, 014002 (2022)

\bibitem{Arrington:2007ux}
J.~Arrington, W.~Melnitchouk, J.A. Tjon, Phys. Rev. C \textbf{76}, 035205
  (2007)

\bibitem{arrington:2011dn}
J.~Arrington, P.G. Blunden, W.~Melnitchouk, Prog. Part. Nucl. Phys.
  \textbf{66}, 782 (2011)

\bibitem{YE20188}
Z.~Ye, J.~Arrington, R.J. Hill, G.~Lee, Phys. Lett. \textbf{B777} (2018)

\bibitem{PhysRevLett.70.718}
A.~Lung, et~al., Phys. Rev. Lett. \textbf{70}, 718 (1993)

\bibitem{PhysRevLett.75.21}
E.E.W. Bruins, et~al., Phys. Rev. Lett. \textbf{75}, 21 (1995)

\bibitem{PhysRevC.48.R5}
P.~Markowitz, et~al., Phys. Rev. C \textbf{48}, R5 (1993)

\bibitem{ANKLIN1994313}
H.~Anklin, et~al., Phys. Lett. \textbf{B336}, 313 (1994)

\bibitem{ANKLIN1998248}
H.~Anklin, et~al., Phys. Lett. \textbf{B428}, 248 (1998)

\bibitem{KUBON200226}
G.~Kubon, et~al., Phys. Lett. \textbf{B524}, 26 (2002)

\bibitem{PhysRevC.75.034003}
B.~Anderson, et~al., Phys. Rev. C \textbf{75}, 034003 (2007)

\bibitem{PhysRevLett.102.192001}
J.~Lachniet, et~al., Phys. Rev. Lett. \textbf{102}, 192001 (2009)

\bibitem{AMROUN1994596}
A.~Amroun, et~al., Nuclear Physics A \textbf{579}, 596 (1994)

\bibitem{6662e8b2e0a649be97d95883e32b1ccd}
D.~Beck, J.~Asai, D.~Skopik, Physical Review C \textbf{25}, 1152 (1982)

\bibitem{PhysRev.138.B57}
H.~Collard, R.~Hofstadter, E.B. Hughes, A.~Johansson, M.R. Yearian, R.B. Day,
  R.T. Wagner, Phys. Rev. \textbf{138}, B57 (1965)

\end{thebibliography}

\end{document}